\documentclass{paper}
\usepackage[latin9]{inputenc}
\usepackage{bbding}
\usepackage{amsmath}
\usepackage{graphicx}

\makeatletter
\newcommand{\lyxaddress}[1]{
\par {\raggedright #1
\vspace{1.4em}
\noindent\par}
}

\usepackage{txfonts}
\usepackage[numbers, sort&compress]{natbib}

\makeatother

\begin{document}

\title{Measurement of spin pumping voltage separated from extrinsic microwave
effects}

\author{Ryo Iguchi$^{1}$\thanks{iguchi@imr.tohoku.ac.jp} and Eiji Saitoh$^{1-5}$}
\maketitle

\lyxaddress{$^{1}$Institute for Materials Research, Tohoku University, Sendai
980-8577, Japan}

\lyxaddress{$^{2}$WPI Advanced Institute for Materials Research, Tohoku University,
Sendai 980-8577, Japan}

\lyxaddress{$^{3}$Center for Spintronics Research Network, Tohoku University,
Sendai 980-8577, Japan}

\lyxaddress{$^{4}$Spin Quantum Rectification Project, ERATO, Japan Science and
Technology Agency, Sendai 980-8577, Japan}

\lyxaddress{$^{5}$Advanced Science Research Center, Japan Atomic Energy Agency,
Tokai 319-1195, Japan}
\begin{abstract}
Conversions between spin and charge currents are core technologies
in recent spintronics. In this article, we provide methods for estimating
inverse spin Hall effects (ISHEs) induced by using microwave-driven
spin pumping (SP) as a spin-current generator. ISHE and SP induce
an electromotive force at the ferromagnetic or spin-wave resonance,
which offers a valuable electric way to study spin physics in materials.
At the resonance, a microwave for exciting the magnetization dynamics
induces an additional electromotive force via rf-current rectification
and thermoelectric effects. We discuss methods to separate the signals
generated from such extrinsic microwave effects by controlling sample
structures and configurations. These methods are helpful in performing
accurate measurements on ISHE induced by SP, enabling quantitative
studies on the conversion between spin and charge currents on various
kinds of materials.
\end{abstract}

\section{Introduction}

Currents of spin angular momentum play a central role in the field
of spintronics. Significant contributions have been made by spin currents,
such as control of magnetizations by spin transfer torque,\citep{PhysRevLett.84.3149,demidov2011control,berger1996emission,ralph2008spin}
transmission of electric signals through insulators,\citep{kajiwara2010transmission,cornelissen2015long,:/content/aip/journal/apl/107/17/10.1063/1.4935074}
thermoelectric conversion,\citep{uchida2008observation,uchida2010spin,jaworski2010observation,kirihara2012spin,7452553}
and electric probing of insulator magnetization.\citep{nakayama2013spin,chen2013theory,althammer2013quantitative}
In order to detect and utilize these spin based-phenomena, conversion
between spin and charge currents is necessary. For realizing efficient
spin-to-charge current conversion, a wide range of materials are currently
being investigated, including metals,\citep{saitoh2006conversion,kimura2007room,miao2013inverse,liu2012spin,mosendz2010quantifying,wang2014scaling,valenzuela2006direct,mendes2014large,niimi2015reciprocal,Azevedo:2005kw}
semiconductors,\citep{wunderlich2005experimental,sinova2004universal,kato2004observation,ando2011electrically,ando2012observation,murakami2003dissipationless,chen2013direct}
organic materials,\citep{SunSchootenKavandEtAl2016,ando2013solution}
carbon-based materials,\citep{dushenko2015experimental} and topological
insulators.\citep{shiomi2014spin,AndoHamasakiKurokawaEtAl2014,mellnik2014spin}
Finding materials suitable to the spin-to-charge conversion is thus
indispensable to making spintronic devices.

One of the popular methods of the spin-to-charge conversion is the
inverse spin Hall effect (ISHE),\citep{Azevedo:2005kw,saitoh2006conversion}
which is the reciprocal effect of the spin Hall effect\citep{hirsch1999spin,kato2004observation,sinova2004universal,murakami2003dissipationless}
caused by spin-orbit interaction. In the ISHE, a spin current generates
a transverse charge current in a conductor such as Pt. Since the first
demonstration of the ISHE in Pt and Al,\citep{saitoh2006conversion,valenzuela2006direct}
it has been extensively studied because of its versatility.\citep{hoffmann2013spin,RevModPhys.87.1213}

Dynamical generation of spin currents can be achieved by the spin
pumping (SP).\citep{mizukami2002effect,Tserkovnyak:2005fr,tserkovnyak2002enhanced}
At the interface between a normal conductor (N) and a ferromagnet
(F), the SP causes emission of spin currents into the N layer from
magnetization dynamics in the adjacent F layer. Such the magnetization
dynamics is typically triggered by applying a microwave field; at
the ferromagnetic resonance (FMR) or the spin wave resonance (SWR)
condition, the magnetization resonantly absorbs the microwave power
and exhibits a coherent precessional motion. A part of the angular
momenta stored in this precessional motion is the source of the spin
current generated by the SP.

The combination of the ISHE and the SP enables electric detection
and generation of spin currents.\citep{saitoh2006conversion,Azevedo:2005kw}
This is the setup commonly used to study the properties of spin-to-charge
current conversion and spin transport in materials. A spin current
is injected into an N layer by the SP and is converted to a measurable
electromotive force by the ISHE. The conversion efficiency between
spin and charge currents in this process can be determined by estimating
the density of the injected spin current by analyzing microwave spectra.\citep{ando2011inverse,mosendz2010quantifying}
The spin transport property of a material can be investigated by constructing
a heterostructure in which the material of interest is placed between
a spin-current injector and detector layers.\citep{shikoh2013spin,dushenko2015experimental,watanabe2014polaron}

\begin{figure}
\begin{centering}
\includegraphics{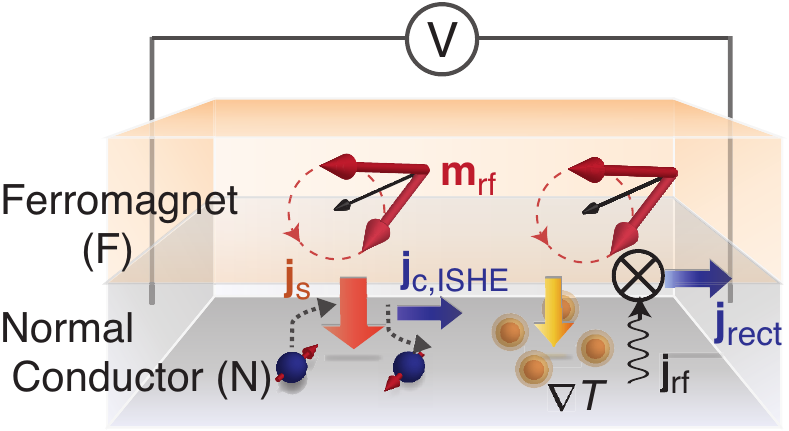}
\par\end{centering}
\caption{(Color online) A schematic illustration of the SP and ISHE processes.
The SP induces an spin current, $\mathbf{j}_{{\rm s}}$, and the ISHE
convert $\mathbf{j}_{{\rm s}}$ into an charge current, $\mathbf{j}_{{\rm c,ISHE}}$.
A microwave driving the FMR induces an rf current, $\mathbf{j}_{{\rm rf}}$,
causing a dc current $j_{{\rm rect}}$ via galvanomagnetic effects.
The absorbed microwave power at FMR induces a temperature gradient,
$\nabla T$, causing a thermoelectric voltage. These process results
in unwanted signals.\label{fig:A-schematic-illustration}}
\end{figure}
It should be noted that the voltage signal from the ISHE can be contaminated
by other contributions in practice experiments. We thus need to extract
the ISHE contribution by separating or minimizing the unwanted signals
in order to ensure the validity of the measurements.\citep{inoue2007detection,mosendz2010quantifying,azevedo2011spin,chen2013direct,iguchi2013effect,PhysRevLett.111.217602}
The ISHE signal is characterized by Lorentzian spectral shape and
sign change under the magnetization reversal,\citep{ando2011inverse}
and some of the unwanted signals show the same spectral shape and
the sign change in configurations commonly used in the SP experiments.\citep{chen2013dc,chen2013direct}
Such signals can be induced by a temperature gradient via thermoelectric
effects\citep{bakker2012thermoelectric,schultheiss2012thermoelectric,shiomi2014spin,qiu2012all}
and by an rf current via rectification effects (See Fig. \ref{fig:A-schematic-illustration}).\citep{Egan:1963gr,1970PhRvB...2.2651M,2007ApPhL..90r2507Y,Gui:2007fb}
The heat emitted by excited magnetization dynamics induces a thermal
gradient in a sample, resulting in an electromotive force due to the
conventional Seebeck effect. This type of heat induced signals can
be eliminated if one designs experimental conditions appropriately.
Rectification effects comes from interplay between stray rf currents
induced by an incident microwave and galvanomagnetic effects coming
from oscillating magnetizations. The direction, magnitude, and phase
of the stray rf current contain uncertainty because they depend on
the details of an experimental setting, so the signals from the rectification
effects tend to be complicated. Since the rectification effects were
first observed in 1963 in a Ni film with the anisotropic magnetoresistance
(AMR) and the anomalous Hall effect (AHE),\citep{Egan:1963gr} we
can not make light of the effects. A number of works regarding to
the extraction of the ISHE signal from other contributions in electric
measurements on the SP have been reported.\citep{inoue2007detection,mosendz2010quantifying,azevedo2011spin,chen2013direct,iguchi2013effect,PhysRevLett.111.217602}

In this article, we review the previous studies of the voltage signals
induced by the SP and further introduce some methods to analyze the
signals on the FMR. Here, we will focus on the experiments in a microwave
cavity, so that small density of induced rf currents and thus small
rectification contribution can be expected. The original field distribution
in a cavity is minimally disturbed by placing a sample in it because
the empty region in the cavity does not carry rf currents. In experiments,
it is often difficult to identify the origin of the stray rf currents
because it depends on an individual setup: the sample structure including
wires for electric measurements. In this paper, we describe methods
to separate the SP contribution from other artifacts in the electric
measurements by introducing parameter dependence to the voltages in
the presence of the stray rf currents. The methods can be applied
to systems with a microstrip line or coplanar waveguide, but the rf-current-induced
magnetization excitation due to the SHE, or the spin transfer torque
FMR (STT-FMR), \citep{chiba2014current,liu2011spin} is neglected.

This article is organized as follows. In Section \ref{sec:signals-due-to},
we show the analytical descriptions of the signals from the SP, rectification
effects, and heating effects based on the magneto-circuit theory\citep{Tserkovnyak:2005fr}
and the Landau-Lifshitz-Gilbert (LLG) equation\citep{gilbert2004phenomenological}.
In Section \ref{sec:Separation-methods}, we discuss the dependence
of the signals on the FMR spectrum, sample geometry, magnetization
orientation, and excitation frequency. Here, we point out the similarity
between the voltage signals due to the SP and rectification effects.
Finally, in Section \ref{sec:Summary}, we give a summary of the methods
based on the dependences described in Section \ref{sec:Separation-methods}
and emphasize that the voltage measurement of the in-plane magnetization
angular dependence with out-of-plane microwave magnetic field in a
properly designed system is the most reliable way and thus enables
quantitative studies of the SP on various kinds of materials.

\section{Signals due to spin pumping and microwave effects in bilayer systems\label{sec:signals-due-to}}

In this section, we discuss the analysis of the voltage signals induced
by the SP, rectification effects, and heating effects. For the calculation
of the SP-induced ISHE signal, we consider a spin current generated
by magnetization dynamics via the SP in a bilayer film consisting
of a normal conductor (N) and a ferromagnet (F). The rectification
signals are expressed using the derived magnetization dynamics for
the SP. For the heating-induced signals, we discuss thermoelectric
effects due to the heating at the FMR and SWR of magnetostatic surface
spin waves.
\begin{figure}
\begin{centering}
\includegraphics{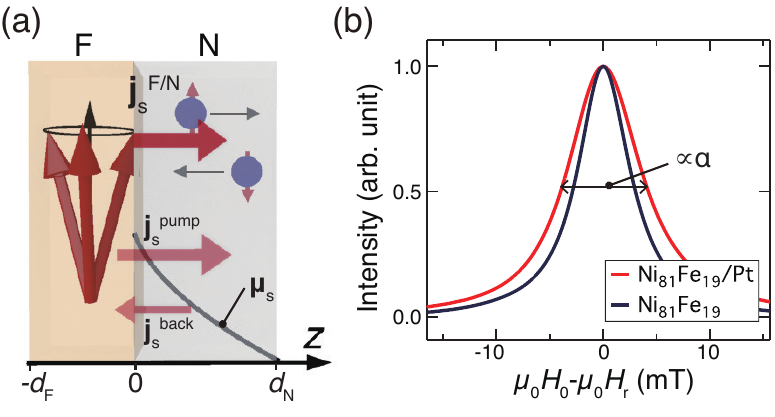}
\par\end{centering}
\caption{(Color online) (a) F/N bilayer system under the SP. F (N) denotes
the ferromagnet (normal conductor) layer. (b) Normalized FMR spectra
of ${\rm Ni}_{81}{\rm Fe}_{19}$ and ${\rm Ni}_{81}{\rm Fe}_{19}$/Pt
systems, where $H_{0}$ denotes the strength of an applied field and
$H_{{\rm r}}$ denotes the FMR field. \label{fig:pumped-and-back}}
\end{figure}

\subsection{Spin current induced by spin pumping}

The spin pumping (SP) is the phenomenon that a magnetically excited
F layer induces a spin current into the N layer placed adjacent to
it.\citep{mizukami2002effect,tserkovnyak2002enhanced} Here, let us
suppose the N and the F layers span $xy$ plane, and they are stacked
in z direction {[}Fig. \ref{fig:pumped-and-back}(a){]}. The spin
current density through the F/N interface due to the SP is given by\citep{Tserkovnyak:2005fr}
\begin{equation}
\mathbf{j}_{{\rm s}}^{{\rm {\rm pump}}}=\frac{\hbar}{4\pi}g_{{\rm r}}^{\uparrow\downarrow}\left(\mathbf{m}\times\dot{\mathbf{m}}\right)+\frac{\hbar}{4\pi}g_{{\rm i}}^{\uparrow\downarrow}\dot{\mathbf{m}},\label{eq:sp}
\end{equation}
where $\mathbf{m}$ denotes the unit vector along the magnetization
in the F layer, $\mathbf{\dot{m}}$ the time derivative of $\mathbf{m}$,
and $g_{{\rm r}}^{\uparrow\downarrow}$ ($g_{{\rm i}}^{\uparrow\downarrow}$)
the real (imaginary) part of mixing conductance per unit area, $\hbar$
the Planck constant. The spin polarization of $\mathbf{j}_{{\rm s}}^{{\rm pump}}$
is represented by its vector direction, and its flow direction is
the interface normal $\boldsymbol{z}$. In diffusive N layers, the
spin accumulation $\boldsymbol{\mu}_{{\rm s}}$ is formed owing to
the pumped spin current. This spin accumulation induces a back-flow
spin current into the F layer, and it renormalizes the mixing conductance
in Eq. (\ref{eq:sp}) to an effective one denoted by $g_{{\rm eff}}^{\uparrow\downarrow}$.
The back-flow spin current density from the magneto-circuit theory
is given by\citep{Tserkovnyak:2005fr} 
\begin{equation}
\mathbf{j}_{{\rm s}}^{{\rm back}}=\frac{1}{4\pi}g_{{\rm r}}^{\uparrow\downarrow}\mathbf{m}\times\left(\boldsymbol{\mu}_{{\rm s}}^{{\rm F/N}}\times\mathbf{m}\right)+\frac{1}{4\pi}g_{{\rm i}}^{\uparrow\downarrow}\boldsymbol{\mu}_{{\rm s}}^{{\rm F/N}}\times\mathbf{m}\label{eq:back}
\end{equation}
when the spin relaxation is fast enough and $\boldsymbol{\mu}_{{\rm s}}^{{\rm F/N}}\propto\mathbf{j_{{\rm s}}^{{\rm pump}}}$
holds,\citep{jiao2013spin} where $\boldsymbol{\mu}_{{\rm s}}^{{\rm F/N}}$
denotes the spin accumulation at the F/N interface ($z=0$). The solution
without the approximation can be found in Ref. \citealp{jiao2013spin}.
Then, combining Eqs. (\ref{eq:sp}) and (\ref{eq:back}), one can
find the net dc spin current density, 
\begin{equation}
\mathbf{j}_{{\rm s}}^{{\rm {\rm F/N}}}=\mathbf{j}_{{\rm s}}^{{\rm pump}}-\mathbf{j}_{{\rm s}}^{{\rm back}}=\frac{\hbar}{4\pi}g_{{\rm r,eff}}^{\uparrow\downarrow}\mathbf{m}\times\dot{\mathbf{m}}+\frac{\hbar}{4\pi}g_{{\rm i,eff}}^{\uparrow\downarrow}\dot{\mathbf{m}},\label{eq:jsdc0}
\end{equation}
where $g_{{\rm r\left(i\right),eff}}^{\uparrow\downarrow}$ represents
the real (imaginary) part of the effective mixing conductance per
unit area. The mixing conductance at the F/N interfaces has been widely
investigated in many combinations of materials.\citep{weiler2013experimental,rojas2014spin,zhang2015role}
Hereafter $g_{{\rm i}{\rm ,eff}}^{\uparrow\downarrow}$ is omitted
for simplicity because it is much smaller than the real part.\citep{Xia:2002ug,Jia:2011gm}

The expression of $g_{{\rm r,eff}}^{\uparrow\downarrow}$ can be obtained
in terms of the parameters of the N layer. $\boldsymbol{\mu}_{{\rm s}}^{{\rm F/N}}$
is calculated from the spin accumulation profile $\boldsymbol{\mu}_{{\rm s}}\left(z\right)$
determined by the spin diffusion equation\citep{jiao2013spin}

\begin{equation}
\frac{\partial}{\partial t}\boldsymbol{\mu}_{{\rm s}}\left(z\right)=-\gamma_{{\rm N}}\boldsymbol{\mu}_{{\rm s}}\left(z\right)\times\mu_{0}\mathbf{H}_{0}+D\nabla^{2}\boldsymbol{\mu}_{{\rm s}}\left(z\right)-\frac{\boldsymbol{\mu}_{{\rm s}}\left(z\right)}{\tau_{{\rm s}}}\label{eq:diffeq}
\end{equation}
with the boundary conditions: $-\frac{\hbar\sigma_{{\rm N}}}{4e^{2}}\nabla\boldsymbol{\mu}_{{\rm s}}\left(0\right)=\mathbf{j}_{{\rm s}}^{{\rm F/N}}$
at the interface ($z=0$) and\linebreak{}
 $-\frac{\hbar\sigma_{{\rm N}}}{4e^{2}}\nabla\boldsymbol{\mu}_{{\rm s}}\left(d_{{\rm N}}\right)=0$
at the outer boundary of the N layer ($z=d_{{\rm N}}$). $\sigma_{{\rm N}\left({\rm F}\right)}$
and $d_{{\rm N}\left({\rm F}\right)}$ are the conductivity and thickness
of the N (F) layer. $\gamma_{{\rm N}}$ denotes the gyromagnetic ratio
of electrons in the N layer, $e$ the electron charge, $\mu_{0}$
the permittivity of a vacuum, $\mathbf{H}_{0}$ an external field,
$D$ the diffusion constant, and $\tau_{{\rm s}}$ the spin-relaxation
time. We focus on the regime where the Hanle effect is negligibly
small (a rigorous treatment can be found in Ref. \citealp{ando2011electrically}).
Then, $\boldsymbol{\mu}_{{\rm s}}\left(z\right)$ is obtained as 
\begin{equation}
\boldsymbol{\mu}_{{\rm s}}\left(z\right)=\frac{4e^{2}}{\hbar\sigma_{{\rm N}}}\lambda\frac{\cosh\left[\left(z-d_{{\rm N}}\right)/\lambda\right]}{\sinh\left(d_{{\rm N}}/\lambda\right)}\mathbf{j}_{{\rm s}}^{{\rm {\rm F/N}}},\label{eq:mus}
\end{equation}
where $\lambda\equiv\sqrt{D\tau_{{\rm s}}}$ denotes the spin diffusion
length. Using Eqs. (\ref{eq:back}), (\ref{eq:jsdc0}), and \eqref{eq:mus},
we find that the effective mixing conductance is given by 
\begin{equation}
g_{{\rm r,eff}}^{\uparrow\downarrow}=\left(\frac{1}{g_{r}}+\frac{\pi\hbar\sigma_{{\rm N}}}{e^{2}\lambda}\text{tanh}\frac{d_{{\rm N}}}{\lambda}\right)^{-1}.\label{eq:geff}
\end{equation}

\subsection{Magnetization dynamics and spin current}

Next, let us examine the effect of the SP on magnetization dynamics.
We will calculate the effective mixing conductance and the magnitude
of a pumped spin current in terms of observable parameters in experiments.
\begin{figure}
\begin{centering}
\includegraphics{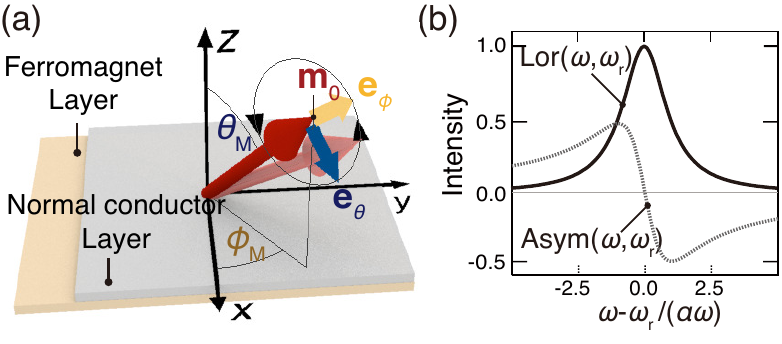}
\par\end{centering}
\caption{(Color online) (a) Spatial coordinate of the system, where $\mathbf{m}_{0}$
denotes unit vector in the direction of the equilibrium magnetization
and is identical with $\mathbf{e}_{r}$. (b) Spectral shapes of ${\rm Lor}\left(\omega,\omega_{{\rm r}}\right)$
and ${\rm Asym}\left(\omega,\omega_{{\rm r}}\right)$. \label{fig:(a)-Spatial-coordinate}}
\end{figure}

The SP affects magnetization dynamics. When the incoming and outgoing
spin currents, given by $\gamma{\rm div}\left({\rm \mathbf{j}_{s}^{{\rm F/N}}}\right)/\left(d_{{\rm F}}I_{{\rm s}}\right)$,
are included, the LLG equation\citep{gilbert2004phenomenological}
is modified as 
\begin{eqnarray}
\dot{\mathbf{m}} & = & -\gamma\mathbf{m}\times\mu_{0}\left[\mathbf{H}_{{\rm eff}}\left(\mathbf{m}\right)+\mathbf{h}_{{\rm rf}}\left(t\right)\right]\nonumber \\
 &  & +\left(\alpha_{0}+\frac{\gamma}{d_{{\rm F}}I_{{\rm s}}}\frac{\hbar}{4\pi}g_{{\rm r,eff}}^{\uparrow\downarrow}\right)\mathbf{m}\times\dot{\mathbf{m}}\label{eq:LLG-1}
\end{eqnarray}
where $\gamma$ denotes the gyromagnetic ratio of the ferromagnet
F, $\mathbf{H}_{{\rm eff}}\left(\mathbf{m}\right)$ the effective
field, $\mathbf{h}_{{\rm rf}}\left(t\right)$ an microwave field,
$\alpha_{0}$ the Gilbert damping constant without spin current exchange,
and $I_{{\rm s}}$ the saturation magnetization. The effective field
is given by $\mathbf{H}_{{\rm eff}}\left(\mathbf{m}\right)=-\nabla_{\mathbf{m}}F_{m}/I_{{\rm s}}$,
where the magnetostatic energy $F_{m}$ includes magnetocrystalline
anisotropy. The second term in the second line of Eq. (\ref{eq:LLG-1})
indicates that the SP acts as an additional damping term.\citep{mizukami2002effect}
The reverse process is also demonstrated; a spin current injected
from the N layer reduces the damping of the F layer.\citep{ando2008electric}

Finding the enhanced damping thus directly relates to $g_{{\rm r,eff}}^{\uparrow\downarrow}$.
In practical experiments, the total damping term represented by $\alpha\mathbf{m}\times\dot{\mathbf{m}}$
is observed. By measuring $\alpha$ from the FMR spectra of F/N and
F films,\citep{tserkovnyak2002spin,mizukami2002effect} the change
in the Gilbert damping constant can be found , i.e. $\Delta\alpha=\alpha_{{\rm F/N}}-\alpha_{{\rm F}}$.
Figure \ref{fig:pumped-and-back}(b) shows an example of the comparison
of the FMR microwave spectrum for ${\rm Ni}_{81}{\rm Fe}_{19}$/Pt
(F/N) bilayer and ${\rm Ni}_{81}{\rm Fe}_{19}$ (F) single-layer samples,
where broadening of the F/N spectral peak can be seen. By fitting
the spectral peak using Lorentzian, $\alpha$ is obtained from the
full-width at the half-maximum (FWHM) which has the relation $\Delta H=\left(\partial\omega_{{\rm r}}/\partial H_{0}\right)^{-1}\alpha\left(\omega_{\theta}+\omega_{\phi}\right)$
for field strength $H_{0}$ swept measurements,\citep{mizukami2002effect}
where $\omega$ and $\omega_{{\rm r}}$ respectively denote the angular
frequency of the magnetization precession and that at the resonance
determined by $\omega_{\theta}$ and $\omega_{\phi}$ derived below.
Finding the value of $\Delta\alpha$, one obtains 
\begin{equation}
g_{{\rm r,eff}}^{\uparrow\downarrow}=4\pi\frac{I_{{\rm s}}}{\gamma\hbar}\Delta\alpha d_{{\rm F}}.\label{eq:geffexp}
\end{equation}
The measurement of $\Delta\alpha$ requires some care. Since the magnetic
properties of a ferromagnetic film in a heterostructure are affected
by the other part of the structure,\citep{johnson1996magnetic} there
can be contributions on $\Delta\alpha$ other than the SP through
inhomogeneous broadening and two-magnon scattering. Thus careful comparison
of $\alpha$ is required.\citep{mizukami2002effect,PhysRevB.73.144424,zakeri2007spin}

In order to find the spectral shape and magnitude of the spin current
$\mathbf{j}_{{\rm s}}^{{\rm F/N}}$, we calculate the magnetization
dynamics $\mathbf{m}\times\dot{\mathbf{m}}$ from Eq. (\ref{eq:LLG-1}).
Here, we consider precessing magnetization $\mathbf{m}_{{\rm rf}}\equiv{\rm Re}\left[\left(m_{\theta}\mathbf{e}_{\theta}+m_{\phi}\mathbf{e}_{\phi}\right)\exp\left(i\omega t\right)\right]$
around the equilibrium magnetization vector $\mathbf{m}_{0}$ excited
by an microwave field $\mathbf{h}_{{\rm rf}}\equiv{\rm Re}\left[\left(h_{\theta}\mathbf{e}_{\theta}+h_{\phi}\mathbf{e}_{\phi}\right)\exp\left(i\omega t\right)\right]$
as shown in Fig. \ref{fig:(a)-Spatial-coordinate}. $\mathbf{m}_{0}$
is determined by the condition $\mathbf{m}_{0}\times\mathbf{H}_{{\rm eff}}=0$.
The unit vectors of the polar coordinate $\mathbf{e}_{p}\left(p=r,\theta,\phi\right)$
in which $\mathbf{e}_{r}$ points $\mathbf{m}_{0}$ have a relation
to the unit vectors of the Cartesian coordinate $\mathbf{e}_{i}\left(i=x,y,z\right)$,
$\mathbf{e}_{p}=\sum_{i=x,y,z}u_{pi}\mathbf{e}_{i}$ with 
\begin{equation}
u=\left(\begin{array}{ccc}
\sin\theta_{{\rm M}}\cos\phi_{{\rm M}} & \sin\theta_{{\rm M}}\sin\phi_{{\rm M}} & \cos\theta_{{\rm M}}\\
\cos\theta_{{\rm M}}\cos\phi_{{\rm M}} & \cos\theta_{{\rm M}}\sin\phi_{{\rm M}} & -\sin\theta_{{\rm M}}\\
-\sin\phi_{{\rm M}} & \cos\phi_{{\rm M}} & 0
\end{array}\right),\label{eq:u}
\end{equation}
where $\theta_{{\rm M}}$ and $\phi_{{\rm M}}$ denote the polar angle
between the $z$ axis and $\mathbf{m}_{0}$, and the azimuthal angle
measured from the $x$ axis, respectively. Taking the time average
of the pumped spin current in Eq. \eqref{eq:jsdc0} yields

\begin{equation}
\left\langle \mathbf{j}_{{\rm s}}^{{\rm F/N}}\right\rangle _{t}=\frac{\hbar\omega}{4\pi}g_{r,{\rm eff}}^{\uparrow\downarrow}{\rm Im}\left[m_{\theta}m_{\phi}^{*}\right]\mathbf{m}_{0},\label{eq:spin pump}
\end{equation}
where $\left\langle \cdots\right\rangle _{t}$ and $a^{*}$ mean temporal
average and complex conjugate of $a$, respectively. The relation
between the magnitude of $\mathbf{m}_{{\rm rf}}$ and $\mathbf{h}_{{\rm rf}}$
obtained from the Eq. \eqref{eq:LLG-1} is reduced to $(m_{\theta},m_{\phi})=\boldsymbol{\chi}.(h_{\theta},h_{\phi})$,
where the susceptibility $\boldsymbol{\chi}$ is given by\citep{smit1959ferrites}
\begin{equation}
\boldsymbol{\chi}\approx\frac{\gamma\mu_{0}}{\alpha\omega}\frac{S\left(\omega,\omega_{{\rm r}}\right)}{\omega_{\theta}+\omega_{\phi}}\left(\begin{array}{cc}
\omega_{\theta} & -\omega_{\theta\phi}+i\omega\\
-\omega_{\theta\phi}-i\omega & \omega_{\phi}
\end{array}\right)\label{eq:chi-1}
\end{equation}
with $\omega_{\theta}\equiv\omega_{\phi\phi}-\frac{\partial}{\partial m_{r}}F_{m}$,
$\omega_{\phi}\equiv\omega_{\theta\theta}-\frac{\partial}{\partial m_{r}}F_{m}$,
and 
\begin{equation}
\omega_{pq}\equiv\frac{\gamma\mu_{0}}{I_{{\rm s}}}\frac{\partial}{\partial m}_{p}\frac{\partial}{\partial m}_{q}F_{m}\ \left(p,q\in\{\theta,\phi\}\right).\label{eq:omegath}
\end{equation}
 $S\left(\omega,\omega_{{\rm r}}\right)$ represents a spectrum function,
\begin{equation}
S\left(\omega,\omega_{{\rm r}}\right)\equiv\frac{\alpha\omega}{\left(\omega_{{\rm r}}-\omega\right)^{2}+\left(\alpha\omega\right)^{2}}\left[\left(\omega_{{\rm r}}-\omega\right)-i\alpha\omega\right],
\end{equation}
which real part represents asymmetric spectrum, ${\rm Asym\left(\omega,\omega_{{\rm r}}\right)\equiv{\rm Re}}\left[S\left(\omega,\omega_{{\rm r}}\right)\right]$
(known as asymmetric Lorentzian), and which imaginary part represents
symmetric spectrum, ${\rm Lor}\left(\omega,\omega_{{\rm r}}\right)\equiv{\rm Im}\left[S\left(\omega,\omega_{{\rm r}}\right)\right]$
(Lorentzian). The resonance angular frequency $\omega_{{\rm r}}$
is given by 
\begin{equation}
\omega_{{\rm r}}\equiv\sqrt{\omega_{\theta}\omega_{\phi}-\omega_{\theta\phi}^{2}}.\label{eq:omegares}
\end{equation}
Calculation using Eq. (\ref{eq:chi-1}) yields 
\begin{equation}
\left\langle \mathbf{j}_{{\rm s}}^{{\rm F/N}}\right\rangle _{t}=\frac{\hbar g_{r,{\rm eff}}^{\uparrow\downarrow}}{4\pi}\frac{\gamma^{2}\mu_{0}^{2}\left(\omega_{\theta}\left|h_{\theta}\right|^{2}+\omega_{\phi}\left|h_{\phi}\right|^{2}+{\rm Im}\left[h_{\theta}h_{\phi}^{*}\left\{ \omega_{\theta}\omega_{\phi}-\left(\omega_{\theta\phi}+i\omega\right)^{2}\right\} /\omega\right]\right)}{\alpha^{2}\left(\omega_{\theta}+\omega_{\phi}\right)^{2}}{\rm Lor}\left(\omega,\omega_{{\rm r}}\right)\mathbf{m}_{0},\label{eq:jsdc}
\end{equation}
where we used a relation, $\left|S\left(\omega,\omega_{{\rm r}}\right)\right|^{2}={\rm Lor}\left(\omega,\omega_{{\rm r}}\right)$.
Equation \eqref{eq:jsdc} means that the spectral shape of $\left\langle \mathbf{j}_{{\rm s}}^{{\rm F/N}}\right\rangle _{t}$
is Lorentzian.\citep{saitoh2006conversion} Equation \eqref{eq:jsdc}
can be also expected in terms of the elliptic precession with the
cone angle $\Theta$ in Refs. \citealp{mosendz2010quantifying,Ando:2009gh,czeschka2011scaling}:
at resonance $\mathbf{m}_{{\rm rf}}={\rm Re}\left[\sin\Theta\exp\left(i\omega t\right)\mathbf{e}_{\theta}+iA\sin\Theta\exp\left(i\omega t\right)\mathbf{e}_{\phi}\right]$
and thus ${\rm Im}\left[m_{\theta}m_{\phi}^{*}\right]=A\sin^{2}\Theta$,
where $A$ is an correction factor for the elliptical precession motion.

The magnetization excitation leads to decreased power of the incident
microwave and thus microwave measurements are useful for determining
$\alpha$ and other parameters related to the resonance condition.
The microwave power absorption on the FMR per unit volume is calculated
by 
\begin{eqnarray}
\Delta P & = & -\frac{\omega}{2\pi}\int_{0}^{\frac{2\pi}{\omega}}\mathbf{h}\cdot\dot{\mathbf{m}}dt\nonumber \\
 & = & \frac{\gamma\mu_{0}}{2}\frac{\left(\omega_{\theta}\left|h_{\theta}\right|^{2}+\omega_{\phi}\left|h_{\phi}\right|^{2}-2{\rm Re}\left[h_{\theta}h_{\phi}^{*}\left(\omega_{\theta\phi}+i\omega\right)\right]\right)}{\alpha\left(\omega_{\theta}+\omega_{\phi}\right)}{\rm Lor}\left(\omega,\omega_{{\rm r}}\right).\label{eq:delP}
\end{eqnarray}
The transferred energy from the microwave to the magnetization dynamics
finally results in heat via damping processes of the dynamics, which
causes thermoelectric signals.

In this calculation, we assumed that $\mathbf{h}_{{\rm rf}}$ is only
induced by an applied microwave field and is not affected by an induced
rf current. Rf currents are known to trigger the FMR through generation
of rf spin currents via the SHE.\citep{liu2011spin} This contribution
can be included to $\mathbf{h}_{{\rm rf}}$ by calculating the spin-transfer
torque due to the absorption of the rf spin current.\citep{liu2011spin,chiba2014current}
Such a contribution will result in a phase shift between the actual
rf current and the rf current determined by the analysis provided
in this article.

\subsection{Voltage generated by inverse spin Hall effect }

Here, we describe the electromotive force generated by the SP and
ISHE in the N layer by taking account of the spin current profile
in the N layer.

The pumped spin current in the N layer gives rise to an electromotive
force due to the ISHE. The ISHE induces a charge current density transverse
to both the spin polarization ($\propto\mathbf{j}_{{\rm s}}$) and
its flow direction $\left(\propto\mathbf{z}\right)$, which can be
expressed as\citep{ando2011inverse} 
\begin{equation}
\mathbf{j}_{{\rm c,ISHE}}=\frac{2e}{\hbar}\theta_{{\rm SHE}}\mathbf{j}_{{\rm s}}\times\mathbf{z}.
\end{equation}
Therefore, as $\left\langle \mathbf{j}_{{\rm s}}^{{\rm F/N}}\right\rangle _{t}\propto\mathbf{m}_{0}$
holds in the SP, the direction of the ISHE current is reversed under
the magnetization reversal ($\mathbf{m}_{{\rm 0}}\rightarrow-\mathbf{m}_{{\rm 0}}$),
which is an important feature of the ISHE induced by the SP.

\begin{figure}
\begin{centering}
\includegraphics{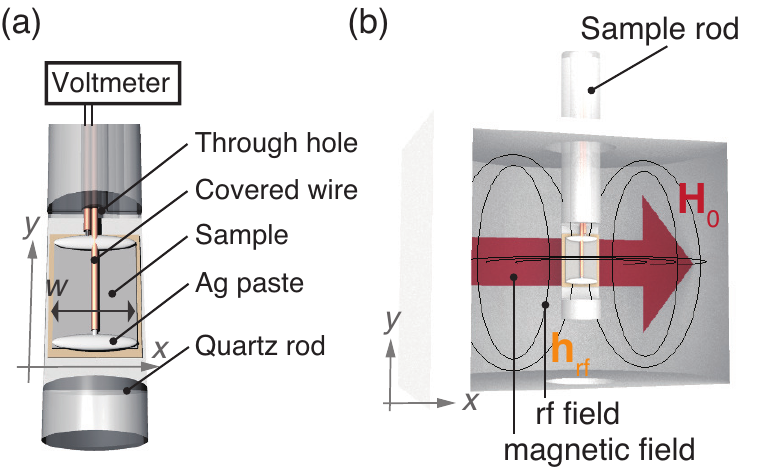}
\par\end{centering}
\caption{(Color online) (a) Typical configuration of sample setup for the ISHE
measurement by using the SP. (b) Schematic illustration of experimental
configuration in a cavity. \label{fig:Typical-configuration-of} }
\end{figure}

According to the short circuit model,\citep{nakayama2012geometry}
the electromotive force is calculated by the sum of the induced current.
The total dc current induced by the ISHE is given by 
\begin{equation}
\mathbf{J}_{{\rm ISHE}}=w\frac{2e}{\hbar}\theta_{{\rm SHE}}\int_{0}^{d_{{\rm N}}}\mathbf{z}\times\left\langle \mathbf{j}_{{\rm s}}\left(z\right)\right\rangle _{t}dz\label{eq:JISHE}
\end{equation}
with $w$ being the width of a sample {[}See Fig. \ref{fig:Typical-configuration-of}(a){]}.
One can observe the electromotive force $\mathbf{E}_{{\rm ISHE}}$
induced by $\mathbf{J}_{{\rm ISHE}}$, which satisfies the relation
$\tilde{R}_{{\rm tot}}\mathbf{J}_{{\rm ISHE}}+\mathbf{E}_{{\rm {\rm ISHE}}}=0$
for an open circuit condition, where $\tilde{R}_{{\rm tot}}$ is the
total resistance per unit length of the system, e.g. $\tilde{R}_{{\rm tot}}^{-1}=w\left(\sigma_{{\rm N}}d_{{\rm N}}+\sigma_{{\rm F}}d_{{\rm F}}\right)$
for F/N bilayer systems. The spin current profile is obtained from
Eqs. (\ref{eq:jsdc0}) and (\ref{eq:mus}) as 
\begin{eqnarray}
\mathbf{j}_{{\rm s}}\left(z\right) & = & -\frac{\hbar}{2e}\frac{\sigma_{{\rm N}}}{2}\nabla\boldsymbol{\mu}_{{\rm s}}\left(z\right)\\
 & = & \mathbf{j}_{{\rm s}}^{{\rm {\rm F/N}}}\frac{\sinh\left(\left[d_{{\rm N}}-z\right]/\lambda\right)}{\sinh\left(d_{{\rm N}}/\lambda\right)}.\label{eq:jsz}
\end{eqnarray}
Then we yields 
\begin{equation}
\mathbf{E}_{{\rm ISHE}}=w\tilde{R}_{{\rm tot}}\theta_{{\rm SHE}}\frac{2e}{\hbar}\lambda\tanh\frac{d_{N}}{2\lambda}\left\langle j_{{\rm s}}^{{\rm {\rm F/N}}}\right\rangle _{t}\mathbf{z}\times\mathbf{m}_{0}\label{eq:EISHE}
\end{equation}
from Eqs. (\ref{eq:jsdc0}), (\ref{eq:JISHE}), and (\ref{eq:jsz}).

A typical experimental setup for observing the ISHE voltage is depicted
in Fig. \ref{fig:Typical-configuration-of}(b). A bilayer sample is
placed on a cutout of a quartz rod, which has a through hole in the
center for wires connected to a voltmeter. The wires are covered by
an insulating polymer and connected to the electrodes in the $y$
axis. These wires measure the $y$ component of Eq. \eqref{eq:EISHE},
so that the maximized ISHE voltage is detected when the magnetization
points along the $x$ axis. $\mathbf{m}_{0}$ is controlled by an
applied field, $\mathbf{H}_{0}$. During the measurements, while $\omega$
is fixed, the field strength $H_{0}$ is swept so that $\omega_{{\rm r}}$
is changed.

\subsection{Voltage generated by rectification effects}

\begin{figure}
\centering{}\includegraphics{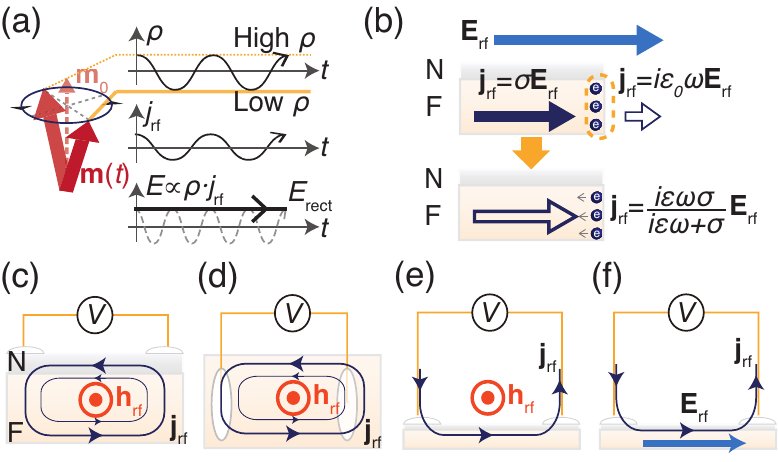}\caption{(Color online) (a) Schematic illustration of mechanism of the rectification
effect. (b) Induced rf current in an isolated conductor under an rf
electric field, $E_{{\rm rf}}$. $\varepsilon_{0}$ denotes the permittivity
of a vacuum (c,d) Possible origins of stray rf currents: induced by
in-plane (c) and out-of-plane (d) rf fields. (e,f) Rf currents picked
up by the wires for the electric measurements. \label{fig:possible-origins-of}}
\end{figure}

Rectification effects at the FMR are caused by an rf current in a
sample possessing a galvanomagnetic effect, i.e. magnetization dependent
resistivity $\rho\left(\mathbf{m}\right)$. Describing the resistivity
which oscillates due to the precessing magnetization as $\tilde{\rho}\left(\mathbf{m}\right)\propto m_{p}\cos\left(\omega t\right)$
and an rf current as $j_{{\rm rf}}=j_{{\rm rf}}^{0}\cos\left(\omega t+\psi\right)$,
a rectified dc electromotive force, $E_{{\rm rect}}$, is given by
\begin{eqnarray}
E_{{\rm rect}} & \propto & m_{p}\cos\left(\omega t\right)\cdot j_{{\rm rf}}^{0}\cos\left(\omega t+\psi\right)\nonumber \\
 & \propto & \frac{jm_{p}}{2}\left[\cos\psi+O\left(t\right)\right],
\end{eqnarray}
where $\psi$ denotes the phase difference between the precessing
magnetization and the rf current, as depicted in Fig. \ref{fig:possible-origins-of}(a).
Here, we will consider the anisotropic magnetoresistance (AMR), the
anomalous Hall effect (AHE), and the spin Hall magnetoresistance (SMR),
which are common examples of galvanomagnetic effects in the bilayer
systems used in the SP measurements. Note that any effect leading
to magnetization dependent resistance gives a dc voltage signal. Therefore
the tunnel magnetoresistance,\citep{yuasa2004giant,ikeda2008tunnel,miyazaki1995giant}
the colossal magnetoresistance,\citep{ramirez1997colossal} and the
spin accumulation Hall effect\citep{hou2015hall} can give the rectified
dc voltages which can be calculated in the way described below.

The induced voltage is calculated from the Ohm's law $E_{i}=\rho_{ij}(\mathbf{m})j_{{\rm rf}}^{j}$,
where $j_{{\rm rf}}^{i}$ denotes the $i$-th component of an rf current
passing through a sample.\citep{Egan:1963gr} For an F layer with
the AMR and AHE, the resistivity tensor can be expanded in terms of
the magnetization by 
\begin{equation}
\rho_{ij}^{{\rm F}}\left(\mathbf{m}\right)=\rho_{ij}^{0}+\Delta\rho_{{\rm AMR}}m_{i}m_{j}+\rho_{{\rm AHE}}\sum_{l=x,y,z}\epsilon_{ijl}m_{l},\label{eq:restensor}
\end{equation}
where $m_{i}$ denotes the component in the $i$ axis and $\epsilon$
denotes the Levi-Civita tensor. $\rho_{ij}^{0}$ denotes the resistivity
part insensitive to $\mathbf{m}$, $\Delta\rho_{{\rm AMR}}$ the AMR
coefficient, and $\rho_{{\rm AHE}}$ the AHE coefficient. For an N
layer with the SMR, whose magnitude is $\Delta\rho_{{\rm SMR}}$,
the resistivity tensor is given by\citep{0953-8984-28-10-103004}
\begin{equation}
\rho_{ij}^{{\rm N}}\left(\mathbf{m}\right)=\rho_{ij}^{0}-\Delta\rho_{{\rm SMR}}\sum_{k=x,y}\epsilon_{ikz}m_{k}\sum_{l=x,y}\epsilon_{jlz}m_{l}+\rho_{{\rm AHE}}^{{\rm N}}\epsilon_{ijz}m_{z}.\label{eq:restensor-1}
\end{equation}
Substituting $m_{i}=\sum_{p=\theta,\phi}u_{pi}m_{p}$ with $u_{pi}$
in Eq. (\ref{eq:u}) and extracting the components proportional to
precessing magnetization, $m_{\theta}$ and $m_{\phi}$, yields the
time-dependent resistivity tensor $\tilde{\rho}^{{\rm F}\left({\rm N}\right)}$
for the F (N) layer {[}cf. Ref. \citealp{iguchi2013effect}{]}. Then
the $i$-th component of the rectified dc current density from the
$j$-th component of the rf current is given by $\sigma_{{\rm F(N)}}{\rm Re}\left[\tilde{\rho}_{ij}^{{\rm F(N)}}j_{{\rm rf}}^{j*}\right]/2$
for the F(N) layer. Considering the short circuit model, the electromotive
force $\mathbf{E}_{{\rm rect}}$ is obtained as 
\begin{equation}
\mathbf{E}_{{\rm rect}}=\tilde{R}_{{\rm tot}}\left(\int_{-d_{{\rm F}}}^{0}\frac{\sigma_{{\rm F}}}{2}{\rm Re}\left[\tilde{\rho}^{{\rm F}}\cdot\mathbf{j}_{{\rm rf}}^{*}\left(z\right)\right]dz+\int_{0}^{d_{{\rm N}}}\frac{\sigma_{{\rm N}}}{2}{\rm Re}\left[\tilde{\rho}^{{\rm N}}\cdot\mathbf{j}_{{\rm rf}}^{*}\left(z\right)\right]dz\right).\label{eq:Erect}
\end{equation}
An origin of $z$ dependence of $\mathbf{j}_{{\rm rf}}\left(z\right)$
comes from the skin effect (which appears as the Dyson effect\citep{dyson1955electron}
for the microwave resonance experiments). $\mathbf{j}_{{\rm rf}}\left(z\right)$
is localized within the skin depth: $\delta_{{\rm skin}}=2\left(\sigma\mu\omega\right)^{-1/2}$,
where $\mu$ denotes the permeability. $\delta_{{\rm skin}}$ is typically
smaller than the thickness of systems used for the SP experiments;
for example, $\delta_{{\rm skin}}$ for Cu at $\omega/2\pi$=10 GHz
is estimated to be 0.6 $\mu$m, and the thickness scale used in the
experiments is less than a few hundred nanometers.
\begin{figure}
\begin{centering}
\includegraphics{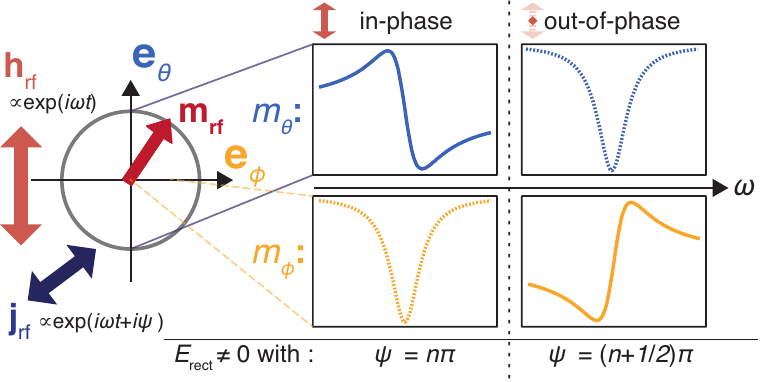}
\par\end{centering}
\caption{(Color online) Spectral shapes of the rectification signal $E_{{\rm rect}}$
induced by $m_{\theta}$ and $m_{\phi}$ at $t=0$ (real part) and
$t=\pi/\left(2\omega\right)$ (imaginary part)\label{fig:(c)-the-spectral}.
$\psi$ is the phase difference between the rf field $h_{{\rm rf}}\exp\left(i\omega t\right)$
and the rf current $j_{{\rm rf}}\exp\left(i\left[\omega t+\psi\right]\right)$. }
\end{figure}

The in-plane electromotive force induced by an rf current which lies
in the film plane ($xy$ plane) and homogeneous over $z$ is 
\begin{align}
\left(\begin{array}{c}
E_{{\rm rect}}^{x}\\
E_{{\rm rect}}^{y}
\end{array}\right)= & \frac{\tilde{R}_{{\rm tot}}}{2}\sin\theta_{{\rm M}}\sum_{L={\rm F},{\rm N}}\sigma_{L}d_{L}\times\nonumber \\
 & \left\{ \left[\cos\theta_{{\rm M}}\rho_{L}^{\theta}+\rho_{{\rm AHE}}^{L}\left(\begin{array}{cc}
0 & 1\\
-1 & 0
\end{array}\right)\right]{\rm Re}\left[m_{\theta}\left(\begin{array}{c}
j_{{\rm rf}}^{x*}\\
j_{{\rm rf}}^{y*}
\end{array}\right)\right]+\rho_{L}^{\phi}{\rm Re}\left[m_{\phi}\left(\begin{array}{c}
j_{{\rm rf}}^{x*}\\
j_{{\rm rf}}^{y*}
\end{array}\right)\right]\right\} \label{eq:Erectex}
\end{align}
with 
\begin{equation}
\rho_{{\rm F(N)}}^{\theta}=\Delta\rho_{{\rm AMR(SMR)}}\left(\begin{array}{cc}
+(-)1+\cos\left(2\phi_{{\rm M}}\right) & \sin\left(2\phi_{{\rm M}}\right)\\
\sin\left(2\phi_{{\rm M}}\right) & +(-)1-\cos\left(2\phi_{{\rm M}}\right)
\end{array}\right),
\end{equation}
\begin{equation}
\rho_{{\rm F(N)}}^{\phi}=\left(\begin{array}{cc}
-\sin\left(2\phi_{{\rm M}}\right) & \cos\left(2\phi_{{\rm M}}\right)\\
\cos\left(2\phi_{{\rm M}}\right) & \sin\left(2\phi_{{\rm M}}\right)
\end{array}\right),
\end{equation}
where + (-) sign on the first term of the diagonal parts corresponds
to the F (N) layer. The magnitude and spectral shape are determined
by ${\rm Re}\left[m_{p}j_{{\rm rf}}^{i{\rm *}}\right]$ through Eq.
(\ref{eq:chi-1}). Figure \ref{fig:(c)-the-spectral} shows possibly
induced spectra with various $j_{{\rm rf}}^{*}$ direction in response
to the dynamic magnetizations, which exhibits both the Lorentzian
and asymmetric Lorentzian. The rectification signals show linear dependence
to the incident power because $m_{p}\propto h_{{\rm rf}}$ and $j_{{\rm rf}}\propto h_{{\rm rf}}$,
which is same as the ISHE signal since $\left\langle j_{{\rm s}}^{{\rm F/N}}\right\rangle _{t}\propto{\rm Im}\left[m_{\theta}m_{\phi}^{*}\right]\propto h_{{\rm rf}}^{2}$.

In the following, we discuss the origin of the stray rf currents causing
the extrinsic signals in the previous experiments. One possible origin
of the rf current in a cavity is the generation due to the Ohm's law
because of a non-zero rf electric field, $\mathbf{E}_{{\rm rf}}$,
which is considered in most of the previous studies.\citep{saitoh2006conversion,ando2011inverse,chen2013direct,inoue2007detection}
However, in an open circuit condition, a conductor smaller than the
microwave wavelength, typically $\sim1$ cm for the X-band microwaves,
can screen $\mathbf{E}_{{\rm rf}}$ below the plasma frequency, which
results in a displacement current $\mathbf{j_{{\rm dis}}}\approx i\varepsilon\omega\mathbf{E}_{{\rm rf}}$
rather than the rf current $\mathbf{j}_{{\rm rf}}=\sigma\mathbf{E}_{{\rm rf}}$
induced by the Ohm's law {[}See Fig. \ref{fig:possible-origins-of}(b){]}.
Here $\varepsilon$ is the permittivity. A simulation in Ref. \citealp{peligrad1998cavity}
shows that the original cavity field distribution is modified so that
$\mathbf{E}_{{\rm rf}}$ in a conductor is zero. $\mathbf{j}_{{\rm dis}}$
is usually much smaller than the $\mathbf{j}_{{\rm rf}}$, e.g. by
a factor of $10^{8}$ for Cu at $\omega/2\pi=10$ GHz, and the phase
of $\mathbf{j}_{{\rm dis}}$ is $90^{\circ}$ different from the $\mathbf{j}_{{\rm rf}}$,
which indicates other consideration is necessary to explain previous
results.

The Faraday's law causes an rf current from a microwave magnetic field.
An in-plane microwave magnetic field induces an eddy current in the
cross section of the bilayer systems {[}See Fig. \ref{fig:possible-origins-of}(c){]}.
Though the current directions are opposite in the F and N layers,
the microwave current can induce a non-zero rectification signal due
to the difference between $\tilde{\rho}^{{\rm F}}\left(\mathbf{m}\right)$
and $\tilde{\rho}^{{\rm N}}\left(\mathbf{m}\right)$. This is one
of plausible contributions to the experiments. For out-of-plane microwave
field, the voltage appears only when the electrodes are placed off
center from the rf-eddy-current distribution {[}See Fig. \ref{fig:possible-origins-of}(d){]}.
Such an effect was studied in Ref. \citealp{EuO}, which remarks that
this effect can be eliminated by making the sample structure symmetric.

Consideration on wires connected to samples for the ISHE measurements
leads to two additional contributions by the Ohm's law and Faraday's
law, which are shown in Fig. \ref{fig:possible-origins-of}(e,f).
In one case, $\mathbf{j}_{{\rm rf}}$ can be proportional to $\sigma\mathbf{E}_{{\rm rf}}$
with a factor considering the modification of the field due to a conducting
sample. $\mathbf{j}_{{\rm rf}}$ is generated and is transmitted through
the paired or twisted wires forming an microwave transmission line.
In the other case, $\mathbf{j}_{{\rm rf}}$ is generated by the induction
around a sample forming a pick-up coil with wires. In both the cases,
rf currents only appear along the electrode direction.

As we discussed above, there are several mechanisms for the stray
rf-current generation in cavities. The suppression of the rf current
is not straightforward and thus the analysis based on $\mathbf{j}_{{\rm rf}}$
by leaving its magnitude, direction and phase undefined parameters,
which are to be fitted with experimental data, is appropriate. The
consideration of in-plane rf currents is enough for the analysis because
the aforementioned mechanisms do not induce an rf current along the
$z$ direction unless a pathway for the rf current is formed in the
$z$ direction. This also holds for rf currents originating from the
magnonic charge pumping,\citep{azevedo2015electrical,ciccarelli2015magnonic}
the ac ISHE current due to the ac SP,\citep{wei2014spin,weiler2014phase}
and other generation effects. 

\subsection{Voltage generated by heating\label{subsec:Induced-voltage-by}}

\begin{figure}
\begin{centering}
\includegraphics{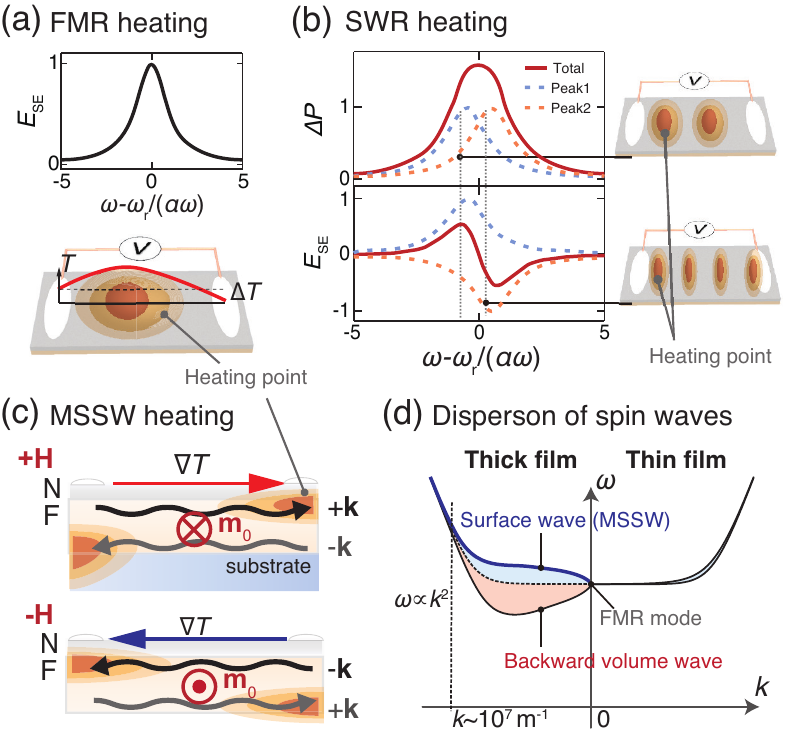}
\par\end{centering}
\caption{(Color online) (a,b) Spectral shape of electromotive forces due to
heat induced by the FMR (a) and SWR (b). (c) Induced $\nabla T$ due
to MSSWs. When the magnetization is reversed, $\nabla T$ can change
its sign, resulting in the same symmetry as the ISHE ($\mathbf{E}_{{\rm ISHE}}\propto\mathbf{z}\times\mathbf{m}_{0}$).
(d) Schematic illustration of difference in the dispersion relation
for thin and thick ferromagnetic films. \label{fig:Electromotive-force-due}}
\end{figure}
Damping processes of magnetization dynamics generate heat, which can
result in an electromotive force via thermoelectric effects. Thermoelectric
effects, such as the Seebeck and Nernst effects, are seen in conducting
materials by themselves regardless of the presence of the F layer.
When a magnetization emits the absorbed power $\Delta P$ to phonons,
a temperature gradient, $\nabla T\propto\Delta P\propto{\rm Lor}\left(\omega,\omega_{{\rm r}}\right)$,
can be induced. $\nabla T$ is then formed into an electromotive force,
e.g. $E_{{\rm SE}}\propto\nabla T$ via the Seebeck effect. While
the spectral shape of $E_{{\rm SE}}$ on the FMR is Lorentzian, it
becomes complicated on the SWR in thick ferromagnetic films, which
is also used to drive the SP.\citep{sandweg2010enhancement} The SWR
gives a different heat profile on each resonance,\citep{An:2013jn}
so that the sign of $\nabla T$ easily changes. When the neighboring
peaks have different signs, the signal should look like the asymmetric
Lorentzian, as shown in Fig. \ref{fig:Electromotive-force-due}(b).
As a result, the total thermoelectric voltage becomes the superposition
of the symmetric and asymmetric Lorentzian for the SWR case.

The Seebeck effect contributes to the voltage signal when the heat
profile produced by the FMR or SWR is not symmetrically distributed
with respect to the electrodes for the voltage detection.In most systems,
it does not have explicit dependence on the magnetization direction.
However, there can be a case where $\nabla T$ is sensitive to the
magnetization direction as is seen in a system comprised of a thick
ferromagnetic film.\citep{an2013unidirectional} In such films, spin
waves known as magnetostatic surface spin waves (MSSWs) localize on
the top and bottom surfaces of the F layer and propagate non-reciprocally.\citep{stancil2009spin}
MSSWs are demonstrated to convey heat to an arbitrary direction controlled
by the magnetization polarization utilizing the non-reciprocity by
unbalanced excitation of spin waves with wavevectors $+\mathbf{k}$
and $-\mathbf{k}$.\citep{an2013unidirectional} This results in $\nabla T\propto\mathbf{k}$
and thus induces a Seebeck voltage in the N layer, $\mathbf{E}_{{\rm SE}\left({\rm MSSW}\right)}\propto\Delta P\mathbf{k}$.
In microstrip antenna excitation, $\mathbf{k}$ of the dominant MSSWs
is reversed under the magnetization reversal, $\mathbf{m}_{0}\rightarrow-\mathbf{m}_{0}$.\citep{an2013unidirectional,schneiderphase}
This is because $\mathbf{k}$ of the MSSWs localized at a surface
with its normal $\mathbf{n}$ is determined by $\mathbf{m}_{0}\times\mathbf{n}$.\citep{stancil2009spin}
Consequently, the sign of the induced heat current by the MSSW is
reversed, resulting in a thermoelectric signal with the same symmetry
as the ISHE, i.e. 
\begin{equation}
\mathbf{E}_{{\rm SE}\left({\rm MSSW}\right)}=A_{{\rm SE\left(MSSW\right)}}\Delta P\mathbf{z}\times\mathbf{m}_{0},\label{eq:Ese}
\end{equation}
where $A_{{\rm SE\left(MSSW\right)}}$ is a constant determined by
the Seebeck coefficient and the temperature profile due to the MSSW.
In the cavity experiments on the SWR, though the spin waves with $+\mathbf{k}$
and $-\mathbf{k}$ are equally excited, asymmetries between the surfaces
that these two modes are localized on can give the thermoelectric
voltage in the same form as Eq. (\ref{eq:Ese}); as we place a N layer
on top of the F layer, the inversion symmetry between the top and
bottom layers is broken, so that the contributions from $+\mathbf{k}$
and $-\mathbf{k}$ give unequal contributions to $\nabla T$ in the
N layer, which results in $\nabla T\propto\mathbf{k}$. Moreover,
the existence of the substrate at the bottom surface may promote the
asymmetry of the thermal conduction, possibly growing $\nabla T$
{[}See Fig. \ref{fig:Electromotive-force-due}(c){]}. Thus, the MSSW
heating effect can appear regardless of the excitation methods in
the F/N bilayer systems. This effect can be significant in materials
with high thermoelectric conversion efficiency, such as low carrier
density conductors.

Other contributions come from the transverse thermoelectric effects
reflecting field or magnetization direction, such as the Nernst-Ettingshausen,
anomalous Nernst effect (ANE) and spin Seebeck effect (SSE). Neglecting
the angular difference between an applied field and $\mathbf{m}_{0}$,
the induced voltage is proportional to $\mathbf{m}_{0}\times\nabla T$.
Thus when $\nabla T$ is formed in the thickness direction, it gives
an in-plane electromotive force 
\begin{equation}
\mathbf{E}_{{\rm TTE}}=A_{{\rm TTE}}\Delta P\mathbf{z}\times\mathbf{m}_{0},
\end{equation}
where $A_{{\rm TTE}}$ denotes a proportionality constant determined
by the magnitude of the transverse thermoelectric effects and $\nabla T$
along $z$ direction. Importantly, $\mathbf{E}_{{\rm TTE}}$ shows
the same symmetry as $\mathbf{E}_{{\rm ISHE}}$.

\section{Separation methods of SP-induced ISHE signal\label{sec:Separation-methods}}

In this section, we will introduce a guideline to select proper materials
for the F layer and four methods to extract the spin current contribution
from observed signals. Here, we discuss the microwave contribution
to the voltage signals in terms of spectral shape, thickness, magnetization
angle, and excitation frequency dependences. Understanding these dependences,
the rectification effects can be isolated by a measurement of magnetization
angular dependence, and the heating effects can be isolated by that
of frequency dependence.

\subsection{Suitable sample design for measurements of ISHE driven by SP\label{subsec:Suitable-materials-for}}

For electric measurements of the SP, an appropriate choice of materials
for reducing the rectification and heating effects can improve the
performance of the experiments. The first step of experiments of the
ISHE induced by the SP thus is to consider the right choice of materials
for the spin injector.

The rectification effects can be suppressed by using a material with
low galvanomagnetic coefficients. The coefficients represented by
$\Delta\rho_{{\rm AMR}}$, $\rho_{{\rm AHE}}^{{\rm F}}$, $\Delta\rho_{{\rm SMR}}$,
and $\rho_{{\rm AHE}}^{{\rm N}}$ in Eq. (\ref{eq:restensor}) are
proportional to the signals. For the metallic spin injector, ${\rm Ni}_{81}{\rm Fe}_{19}$,
so called Permalloy, is often used but other materials such as CoFe
alloys with the low AMR ratio is a good candidate for the SP.\citep{haidar2015reducing,ganguly2014thickness}
Similarly, the SMR is known to be small compared to the AMR, and thus
the use of a ferrimagnetic insulator is effective.\citep{iguchi2013effect}

The FMR and MSSW heating effects due to the conventional Seebeck effect
can be minimized by making the sample structure symmetric about the
electrodes and by reducing the thickness of the ferromagnetic layer
$d_{{\rm F}}$, respectively. The feature due to magnetostatic interaction
is dominant around $\left|\mathbf{k}\right|d_{{\rm F}}\sim1$. \citep{stancil2009spin}
When $d_{{\rm F}}$ is decreased, such a value of $\mathbf{k}$ increases
and eventually reaches the exchange regime where the magnetostatic
feature is lost. Figure \ref{fig:Electromotive-force-due} illustrates
the dispersion relation of spin waves for thick and thin ferromagnetic
films. The manifold of the dispersion shrinks as the film thickness
reduces. The group velocity of the MSSWs correspondingly becomes smaller,\citep{iguchi2013spin}
and the heat conveyer effect eventually disappears. Depending on the
strength of the Seebeck effect, the appropriate thickness is below
$100$ nm for the measurements free from the MSSWs heating, which
can be fabricated by pulsed laser deposition,\citep{PhysRevLett.107.066604}
sputtering,\citep{:/content/aip/journal/apl/101/15/10.1063/1.4759039}
or metal-organic decomposition.\citep{ishibashi2010magneto} The MSSW
contribution can also be confirmed by a control experiment with the
insertion of a thin nonmagnetic insulator layer between the N and
F layers because the nonmagnetic insulator cuts the spin transport
but allows heat transport.\citep{PhysRevLett.111.247202}

\subsection{Spectral shape dependence }

First, let us introduce a way to separate a measured electric signal
into symmetric and antisymmetric parts with respect to reversal of
magnetization. We will explain why this simple method does not work
for isolating the microwave effects. In addition to the spectral shape
separation introduced here, measurements on the other dependences
are strongly recommended.
\begin{figure}
\begin{centering}
\includegraphics{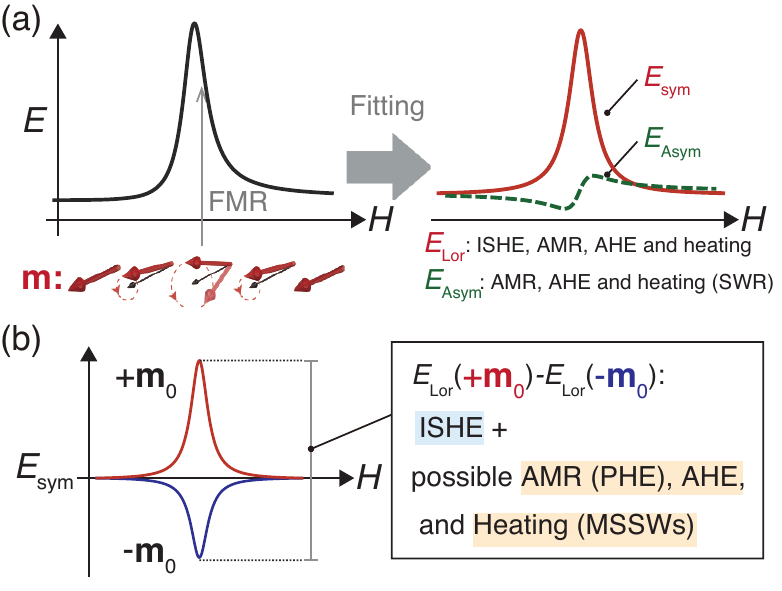}
\par\end{centering}
\caption{(Color online) (a) Common procedure for extracting the ISHE contribution.
(b) The origins which can show the sign change under the magnetization
reversal. Lorentzian part with the sign change is not only due to
the ISHE, but also due to the microwave extrinsic effects. \label{fig:Analysis-of-obtained}}
\end{figure}

As it is derived in Eqs. (\ref{eq:EISHE}) and (\ref{eq:Erect}),
the dc electromotive force on the FMR $\mathbf{E}_{{\rm tot}}=\mathbf{E}_{{\rm ISHE}}+\mathbf{E}_{{\rm rect}}+\mathbf{E}_{{\rm SE}{\rm \left(MSSW\right)}}+\mathbf{E}_{{\rm TTE}}$
has the two distinct parts proportional to ${\rm Lor}\left(\omega,\omega_{r}\right)$
and ${\rm Asym}\left(\omega,\omega_{r}\right)$. By fitting an observed
signal by

\begin{equation}
\mathbf{E}_{{\rm tot}}=\mathbf{E}_{{\rm sym}}{\rm Lor}\left(\omega,\omega_{{\rm r}}\right)+\mathbf{E}_{{\rm asym}}{\rm Asym}\left(\omega,\omega_{{\rm r}}\right),
\end{equation}
the separation can be done, where $E_{{\rm sym}\left({\rm asym}\right)}$
is the magnitude of the ${\rm Lor}\left(\omega,\omega_{{\rm r}}\right)$\linebreak{}
 (${\rm Asym}\left(\omega,\omega_{{\rm r}}\right)$) part (See Fig.
\ref{fig:Analysis-of-obtained}). The earlier naive discussions attribute
the whole $E_{{\rm sym}}$ due to the ISHE, but this assumption does
not hold as is discussed below. Generally, $E_{{\rm sym}}$ includes
not only the ISHE component but also the rectification contribution.\citep{azevedo2011spin,chen2013direct}
For example, in a ${\rm TE}_{011}$ cavity with the rf field $h_{{\rm rf}}$
in the $y$ direction and an rf current along the $x$ direction with
$j_{{\rm rf}}\propto ih_{{\rm rf}}$, the $y$-component of Eq. \eqref{eq:Erectex}
leads to 
\begin{align}
E_{{\rm sym},{\rm ISHE}}^{y}\propto & \frac{\omega_{\phi}}{\left(\omega_{\theta}+\omega_{\phi}\right)^{2}}\sin\theta_{{\rm M}},\nonumber \\
E_{{\rm sym},{\rm rect}}^{y}\propto & \frac{\left(\rho_{{\rm AHE}}\omega_{\theta\phi}-\Delta\rho_{{\rm AMR}}\omega_{\phi}\right)}{\omega\left(\omega_{\theta}+\omega_{\phi}\right)}\sin\theta_{{\rm M}}\label{eq:ESMRy}
\end{align}
at $\phi_{{\rm M}}=0$.\citep{chen2013direct} The configuration is
depicted in Fig. \ref{fig:Typical-configuration-of}(b). $E_{{\rm sym}}$
includes signals due to the rectification effects and, importantly,
possesses the same symmetry as the ISHE signal, i.e. $\sin\theta_{{\rm M}}$.
Therefore, the part of the Lorentzian signal with the sign change
following to the magnetization reversal cannot be attributed to only
the ISHE without further examinations. In Ref. \citealp{chen2013direct},
a separation of the contributions based on the difference in the pre-factors,
i.e. $\omega_{\theta\phi}$, $\omega_{\phi}$, $\omega$, darling
the $\theta_{{\rm M}}$ scan is suggested and will be introduced in
Sect. \ref{subsec:Angular-dependence-of}. Note that Eq. (\ref{eq:ESMRy})
holds only when the rf current is constant during the scan. However,
the stray rf current often shows angular dependence.\citep{lustikova2015vector}

The heating effect due to transverse thermoelectric effects can also
induce the similar signal to the ISHE, which is given by 
\begin{equation}
E_{{\rm sym,T{\rm TE}}}^{y}\propto\frac{\omega_{\phi}}{\omega_{\theta}+\omega_{\phi}}{\rm sin\theta_{{\rm M}}.}
\end{equation}
The FMR heating contribution discussed in Sect. \ref{subsec:Induced-voltage-by}
can be extracted by the frequency dependence measurement. The MSSW
heating contribution discussed in Sect. \ref{subsec:Induced-voltage-by}
is unable to be removed until the F layer thickness is reduced. One
possible solution for handling this difficulty is finding this contribution
from the the calculation based on the Seebeck coefficient following
a temperature profile measurement as is done in Ref. \citealp{qiu2015spin}.

An asymmetric component is a sign of contribution from the rectification
effects although the reverse is not true because rectification signals
can have only the Lorentzian component. If one knows the direction
of the rf current, then the asymmetric component might be a good measure
to determine its magnitude and thus the rectification contribution.
The direction may be estimated by measuring the voltage along other
directions as shown in Ref. \citealp{lustikova2015vector,tsukahara2014self}.

\subsection{Thickness dependence\label{subsec:Thickness-dependence-of}}

\begin{figure}
\begin{centering}
\includegraphics{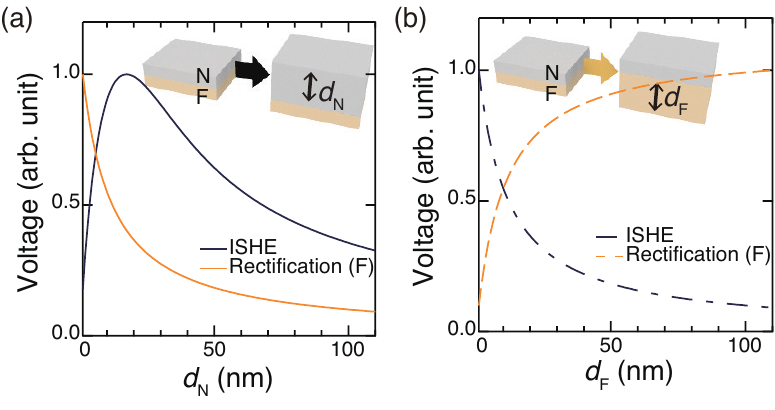}
\par\end{centering}
\caption{(Color online) (a,b) Thickness $d_{{\rm N}}$ (a) and $d_{{\rm F}}$
(b) dependence of the ISHE and rectification signals. We assume $\lambda=10$
nm and normalize the curves by the maximum value in the plot regime.
\label{fig:Thickness-dependence-of}}
\end{figure}
The voltage signals from the ISHE and rectification effects have different
dependence on the thickness of the F and N layers.\citep{nakayama2012geometry}
For a bilayer where the galvanomagnetic effects in the F layer is
dominant, the symmetric Lorentzian signal after taking the difference
between $\mathbf{m}_{0}$ and $-\mathbf{m}_{0}$ is expressed in the
following form, 
\begin{equation}
E_{{\rm sym},{\rm ISHE}}^{y}=\frac{E_{{\rm ISHE}}^{0}}{\sigma_{{\rm N}}d_{{\rm N}}+\sigma_{{\rm F}}d_{{\rm F}}}\tanh\frac{d_{{\rm N}}}{2\lambda},\label{eq:thEyishe}
\end{equation}
\begin{equation}
E_{{\rm sym},{\rm rect}}^{y}=\frac{d_{{\rm F}}E_{{\rm rect}}^{{\rm 0}}}{\sigma_{{\rm N}}d_{{\rm N}}+\sigma_{{\rm F}}d_{{\rm F}}},
\end{equation}
where $E_{{\rm ISHE}}^{0}$ and $E_{{\rm rect}}^{0}$ are respectively
determined by Eqs. (\ref{eq:EISHE}) and (\ref{eq:Erectex}). Regarding
$E_{{\rm ISHE}}^{0}$ (or $\left\langle j_{{\rm s}}^{{\rm F/N}}\right\rangle _{t}$)
and $E_{{\rm rect}}^{0}$ as constants, the $d_{{\rm N}}$ dependence
of Eq. (\ref{eq:thEyishe}) reads a tangent hyperbolic function divided
by the total resistance, which shows a positive peak around $d_{{\rm N}}\approx2\lambda$
{[}See Fig. \ref{fig:Thickness-dependence-of}(a){]}. The $d_{{\rm N}}$
dependence of the rectification contribution shows monotonic decrease
as $d_{{\rm N}}$ increases. The distinction between these two becomes
difficult for $d_{{\rm N}}>\lambda_{{\rm N}}$, because both show
just decrease as $d_{{\rm N}}$ increases {[}See Fig. \ref{fig:Thickness-dependence-of}(b){]}.
On the other hand, the $d_{{\rm F}}$ dependence shows clear difference;
as $d_{{\rm F}}$ increases, the ISHE and rectification contributions
respectively show monotonic decrease and increase before saturation.
This is because the SP (the rectification effects) induced voltage
is electrically-shorted by the additional F (N) layer. This contrary
dependence of the SP and rectification contributions is very useful
to separate them. For a practice analysis, thickness dependence of
parameters should be considered, such as $g_{{\rm eff,r}}^{\uparrow\downarrow}$,
$\tilde{\rho}^{{\rm F}}$, and $j_{{\rm rf}}\left(z\right)$.

The FMR and MSSW heating effects can complicate the thickness dependence
of the signal because the induced temperature gradient depends on
sample structure and environment. Therefore, it is highly recommended
to carefully consider an appropriate design of the samples as described
in Sect. \ref{subsec:Suitable-materials-for}.

\subsection{Angular dependence\label{subsec:Angular-dependence-of}}

\begin{figure}[t]
\begin{centering}
\includegraphics{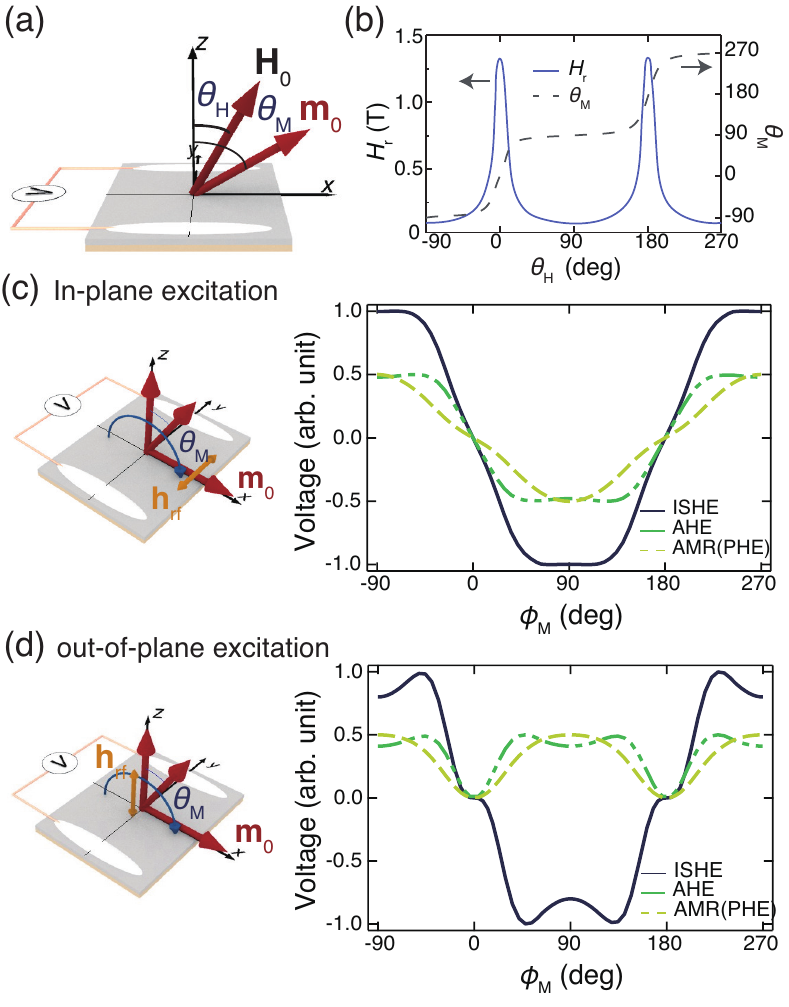}
\par\end{centering}
\caption{(Color online) (a) Schematic illustration of out-of-plane (OP) angular
dependence measurement, where $\mathbf{m}_{0}$ and $\mathbf{H}_{0}$
points different direction due to the demagnetizing and uniaxial anisotropy
fields. (b) $H_{{\rm r}}$ and $\theta_{{\rm M}}$ as a function of
$\theta_{{\rm H}}$. (c,d) angular dependence of the ISHE and rectification
signals for the in-plane excitation (c) and OP excitation (d). The
ISHE and rectification signals are normalized to be 1 and 0.5, respectively.\label{fig:OP-angular-dependence}}
\end{figure}

\begin{figure}
\begin{centering}
\includegraphics{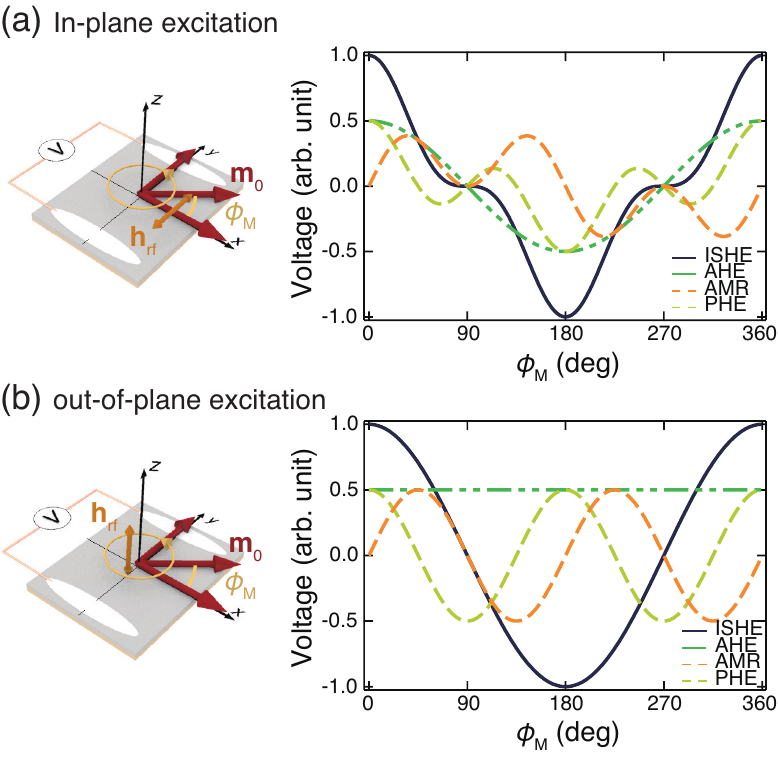}
\par\end{centering}
\caption{(Color online) (a, b) In-plane (IP) angular dependence of the ISHE
and rectification signals for the IP excitation (a) and out-of-plane
excitation (b). The ISHE and rectification signals are normalized
to be 1 and 0.5, respectively. PHE denotes the AMR signal caused by
$j_{{\rm rf}}^{x}$.\label{fig:IP-angular-dependence}}
\end{figure}

The magnetization angular dependence of $\mathbf{E}_{{\rm ISHE}}$
is derived in Ref. \citealp{ando2008angular} and that of $\mathbf{E}_{{\rm rect}}$
due to the AMR (PHE), AHE and SMR can be found in Refs. \citealp{chen2013direct,iguchi2013effect,Egan:1963gr,mecking2007microwave,yamaguchi2008broadband}.
Here, we will focus on the ISHE and rectification effects, because
the heating effects can not be isolated by the angular dependence.
The angular dependence can be studied by two types of rotation: out-of-plane
angular dependence (OP) and in-plane angular dependence (IP), which
are shown in Figs. \ref{fig:OP-angular-dependence} and \ref{fig:IP-angular-dependence}.
In addition, there are two different means to excite the ferromagnet:
applying an rf excitation field in OP or IP. Here, these four configurations
are considered in a film system with uniaxial anisotropy perpendicular
to the film plane, which is relevant to most of experiments. Such
systems are described by the magnetostatic energy: 
\begin{equation}
F_{m}=-\mathbf{m}\cdot\mathbf{H}_{0}+\frac{I_{{\rm s}}^{2}}{2\mu_{0}}m_{z}^{2}-K_{{\rm u}}m_{z}^{2},\label{eq:Um}
\end{equation}
where $\mathbf{H}_{0}$ is an applied external field, given by\linebreak{}
 $\mathbf{H}_{0}=H_{0}\left(\sin\theta_{{\rm H}}\cos\phi_{{\rm H}}\mathbf{e}_{x}+\sin\theta_{{\rm H}}\sin\phi_{{\rm H}}\mathbf{e}_{y}+\cos\theta_{{\rm H}}\mathbf{e}_{z}\right)$,
and $K_{{\rm u}}$ denotes the perpendicular anisotropy constant.
Equation (\ref{eq:Um}) reduces Eq. (\ref{eq:omegath}) to 
\begin{align}
\omega_{\theta} & =\gamma\left[\mu_{0}H_{0}\cos\left(\theta_{{\rm H}}-\theta_{{\rm M}}\right)-I_{{\rm eff}}\cos^{2}\theta_{{\rm M}}\right],\\
\omega_{\phi} & =\gamma\left[\mu_{0}H_{0}\cos\left(\theta_{{\rm H}}-\theta_{{\rm M}}\right)-I_{{\rm eff}}\cos\left(2\theta_{{\rm M}}\right)\right],
\end{align}
and $\omega_{\phi\theta}=0$, where $I_{{\rm eff}}$ denotes the effective
magnetization $I_{{\rm eff}}=I_{{\rm s}}-2\mu_{0}K_{{\rm u}}/I_{{\rm s}}$,
and $\theta_{{\rm H}}$ $\left(\phi_{{\rm H}}\right)$ denotes the
polar (azimuthal) angle in the spherical coordinate as shown in Fig.
\ref{fig:OP-angular-dependence}(a). The resonance field $H_{{\rm r}}$
is determined by $H_{0}$ which simultaneously satisfies $\omega_{{\rm r}}=\sqrt{\omega_{\theta}\omega_{\phi}}$
and $\mathbf{m}_{0}\times\mathbf{H}_{{\rm eff}}=0$, which is reduced
to 
\begin{equation}
2\mu_{0}H_{0}\sin\left(\theta_{{\rm H}}-\theta_{{\rm M}}\right)+I_{{\rm eff}}\sin\left(2\theta_{{\rm M}}\right)=0
\end{equation}
and $\phi_{{\rm H}}=\phi_{{\rm M}}$. Note that in the following calculation
we still use $\omega_{{\rm \theta}\left(\phi\right)}$ for simple
notation but impose $\omega_{{\rm \theta\phi}}=0$.

In the setup for the OP angular dependence measurements, the magnetization
is rotated in the $xz$ plane ($\phi_{{\rm M}}=0$) and electrodes
for detecting the ISHE voltage are placed in the $y$ axis {[}See
Fig. \ref{fig:OP-angular-dependence}(a){]}. An IP excitation field,
$h_{{\rm rf}}$ along the $y$ axis, induces 
\begin{equation}
E_{{\rm sym},{\rm ISHE}}^{y}=A_{{\rm ISHE}}\frac{\omega_{\phi}}{\left(\omega_{\theta}+\omega_{\phi}\right)^{2}}\sin\theta_{{\rm M}},
\end{equation}
\begin{equation}
E_{{\rm sym},{\rm rect}}^{y}=A_{{\rm rect}}\frac{\left(\omega\rho_{{\rm AHE}}{\rm Re}\left[j_{{\rm rf}}^{x*}\right]+\omega_{\phi}\Delta\rho_{{\rm AMR}}{\rm Im}\left[j_{{\rm rf}}^{x*}\right]\right)}{\omega\left(\omega_{\theta}+\omega_{\phi}\right)}\sin\theta_{{\rm M}},
\end{equation}
where the coefficients $A$ are given by 
\begin{equation}
A_{{\rm ISHE}}=w\tilde{R}_{{\rm tot}}\theta_{{\rm SHE}}\frac{2e}{\hbar}\lambda\tanh\frac{d_{{\rm N}}}{2\lambda}\frac{\hbar g_{r,{\rm eff}}^{\uparrow\downarrow}}{4\pi}\frac{\gamma^{2}\mu_{0}^{2}h_{{\rm rf}}^{2}}{\alpha^{2}},
\end{equation}
\begin{equation}
A_{{\rm rect}}=\tilde{R}_{{\rm tot}}\frac{\sigma_{{\rm F}}d_{{\rm F}}}{2}\frac{\gamma\mu_{0}h_{{\rm rf}}}{\alpha},
\end{equation}
both of which are proportional to $\sin\theta_{{\rm M}}$. In Ref.
\citealp{chen2013direct}, the separation is done based on the difference
in the dependence on $\omega_{\theta}\left(\theta_{{\rm M}}\right)$
and $\omega_{\phi}\left(\theta_{{\rm M}}\right)$ between the ISHE
and rectification voltages, which is because they have different responses
to the effective field sweeping over $\theta_{{\rm H}}$. Figure \ref{fig:OP-angular-dependence}(b)
shows $H_{{\rm r}}$ and $\theta_{{\rm M}}$ as a function of $\theta_{{\rm H}}$
calculated with $\gamma=1.79\times10^{11}$ ${\rm T^{-1}s^{-1}}$
and $I_{{\rm eff}}=1$ T. Figure \ref{fig:OP-angular-dependence}(c)
shows the $\theta_{{\rm M}}$ dependence of $E_{{\rm ISHE}}$ and
$E_{{\rm rect}}$, which possess similar form, but slight difference
seen in the solid and dashed curves. This similarity can be a large
source of error in data fitting.

An OP excitation field, $h_{z}$, induces 
\begin{equation}
E_{{\rm sym},{\rm ISHE}}^{y}=A_{{\rm ISHE}}\frac{\omega_{\theta}}{\left(\omega_{\theta}+\omega_{\phi}\right)^{2}}\sin^{3}\theta_{{\rm M}},
\end{equation}
\begin{equation}
E_{{\rm sym},{\rm rect}}^{y}=A_{{\rm rect}}\frac{\left(\omega\Delta\rho_{{\rm AMR}}{\rm Re}\left[j_{{\rm rf}}^{x*}\right]-\omega_{\theta}\rho_{{\rm AHE}}{\rm Im}\left[j_{{\rm rf}}^{x*}\right]\right)}{\omega\left(\omega_{\theta}+\omega_{\phi}\right)}\sin^{2}\theta_{{\rm M}},
\end{equation}
{[}See Fig. \ref{fig:OP-angular-dependence}(d){]}. At a glance, the
difference in the two signals are clear, namely $E_{{\rm rect}}^{y}\propto\sin^{2}\theta_{{\rm M}}$
and $E_{{\rm ISHE}}\propto\sin^{3}\theta_{{\rm M}}$. However, when
$j_{{\rm rf}}\propto\sin\theta_{{\rm M}}$ holds, the contributions
cannot be separated by the harmonic functions.

Similarly to the OP angular dependence in the OP excitation, the IP
angular dependence excited by an IP rf field has the same difficulty.
In the setup, the magnetization is rotated in the $xy$ plane ($\theta_{{\rm M}}=90^{\circ}$)
and electrodes for detecting the ISHE voltage are placed in the $y$
axis. The electromotive forces in the $y$ direction are 
\begin{equation}
E_{{\rm sym},{\rm ISHE}}^{y}=A_{{\rm ISHE}}\frac{\omega_{\phi}}{\omega\left(\omega_{\theta}+\omega_{\phi}\right)^{2}}\cos^{3}\phi_{{\rm M}},
\end{equation}
\begin{equation}
E_{{\rm sym},{\rm rect}}^{y}=A_{{\rm rect}}\cos\phi_{{\rm M}}\frac{\omega_{\phi}\Delta\rho_{{\rm AMR}}\left(\cos\left(2\phi_{{\rm M}}\right){\rm Im}\left[j_{{\rm rf}}^{x*}\right]+\sin\left(2\phi_{{\rm M}}\right){\rm Im}\left[j_{{\rm rf}}^{y*}\right]\right)+\omega\rho_{{\rm AHE}}{\rm Re}\left[j_{{\rm rf}}^{x*}\right]}{\omega\left(\omega_{\theta}+\omega_{\phi}\right)}\label{eq:EsymIPIP}
\end{equation}
{[}See Fig. \ref{fig:IP-angular-dependence}(a){]}. In this configuration,
the ISHE and AMR contributions mix even for the simplest condition
that the rf current is angular-independent.

The IP angular dependence excited by an OP rf field has an advantage
to the previous three angular dependences because the AMR and AHE
show different symmetric angular dependences in the IP configuration.
An OP excitation field, $h_{z}$, induces 
\begin{equation}
E_{{\rm sym},{\rm ISHE}}^{y}=A_{{\rm ISHE}}\frac{\omega_{\theta}}{\left(\omega_{\theta}+\omega_{\phi}\right)^{2}}\cos\phi_{{\rm M}},\label{eq:IP-1}
\end{equation}
\begin{equation}
E_{{\rm sym},{\rm rect}}^{y}=A_{{\rm rect}}\frac{\omega\Delta\rho_{{\rm AMR}}\left(\cos\left(2\phi_{{\rm M}}\right){\rm Re}\left[j_{{\rm rf}}^{x*}\right]+\sin\left(2\phi_{{\rm M}}\right){\rm Re}\left[j_{{\rm rf}}^{y*}\right]\right)-\omega_{\theta}\rho_{{\rm AHE}}{\rm Im}\left[j_{{\rm rf}}^{x*}\right]}{\omega\left(\omega_{\theta}+\omega_{\phi}\right)}.\label{eq:IP-2}
\end{equation}
For the simplest case where $j_{x\left(y\right)}^{{\rm rf}}$ is constant
during the rotation, importantly, the ISHE, AMR and AHE show the different
angular dependences, $\cos\phi_{{\rm M}}$, $\cos\left(2\phi_{{\rm M}}\right)$,
and constant {[}Fig. \ref{fig:IP-angular-dependence}(b){]}. Thus,
fitting the result using the harmonic functions gives the ISHE contribution
directly. Figure \ref{fig:IP-angular-dependence} shows the calculation
result based on Eqs. (\ref{eq:IP-1}) and (\ref{eq:IP-2}).
\begin{figure}
\begin{centering}
\includegraphics{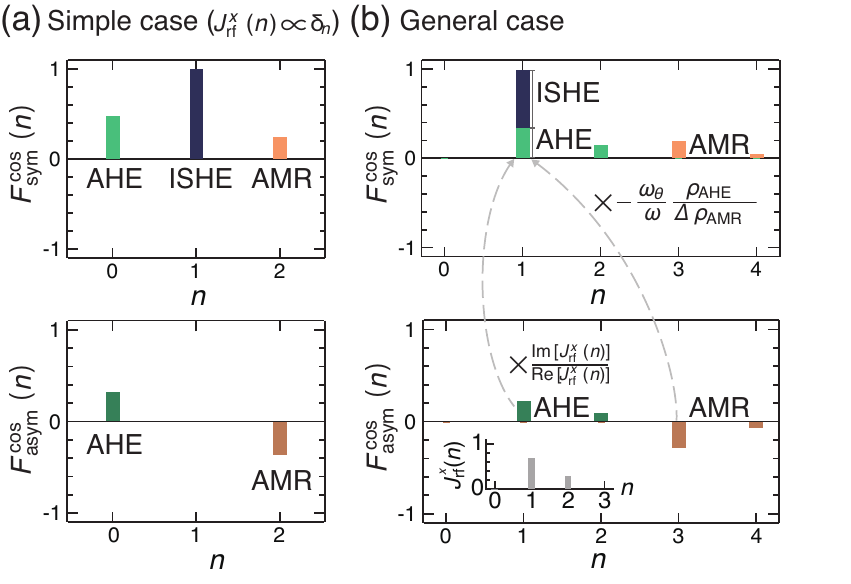}
\par\end{centering}
\caption{(Color online) (a,b) Analysis scheme based on the Fourier cosine coefficients:
case for the constant rf current (a) and case for angular dependent
rf current (b) during the magnetization angle scan in the film plane.
The inset shows the magnitude of the Fourier cosine coefficient of
the complex $j_{{\rm rf}}^{x}$. \label{fig:Analysis-scheme-based}}
\end{figure}

When $j_{x\left(y\right)}^{{\rm rf}}$ has an angular dependence,
an analysis method based on the Fourier series coefficient is effective
in the measurements on the IP angular dependence with OP excitation.
The $n$-th Fourier cosine coefficient of the voltage is given by
\begin{align}
F_{{\rm sym}}^{{\rm cos}}\left(n\right) & =\frac{\omega_{\theta}A{}_{{\rm ISHE}}}{\left(\omega_{\theta}+\omega_{\phi}\right)^{2}}\delta_{n-1}+\frac{A_{{\rm rect}}}{\omega\left(\omega_{\theta}+\omega_{\phi}\right)}\left(\omega\Delta\rho_{{\rm AMR}}{\rm Re}\left[J_{{\rm rf}}^{x}\left(n-2\right)\right]-\omega_{\theta}\rho_{{\rm AHE}}{\rm Im}\left[J_{{\rm rf}}^{x}\left(n\right)\right]\right),\\
F_{{\rm asym}}^{{\rm cos}}\left(n\right) & =\frac{A_{{\rm rect}}}{\omega\left(\omega_{\theta}+\omega_{\phi}\right)}\left(-\omega\Delta\rho_{{\rm AMR}}{\rm Im}\left[J_{{\rm rf}}^{x}\left(n-2\right)\right]-\omega_{\theta}\rho_{{\rm AHE}}{\rm Re}\left[J_{{\rm rf}}^{x}\left(n\right)\right]\right),\label{eq:Fsymasym}
\end{align}
where $\delta_{n}$ denotes the Kronecker delta function and $J_{{\rm rf}}^{x}\left(n\right)$
is the $n$-th Fourier cosine coefficient of $j_{{\rm rf}}^{x}$.
Figure \ref{fig:Analysis-scheme-based}(a) shows an expected intensity
of the coefficients for $J_{{\rm rf}}^{x}\left(n\right)\propto\delta_{n}$.
The contributions are clearly separated. Figure \ref{fig:Analysis-scheme-based}(b)
shows a calculation of the coefficients for $J_{{\rm rf}}^{x}\left(n\right)$
with $J_{{\rm rf}}^{x}\left(1\right)\neq0$. $J_{{\rm rf}}^{x}\left(1\right)$
can induces an ISHE-like signal via the AHE. However, this contribution
can be removed by comparing symmetric and antisymmetric components
because there is a relation, 
\[
F_{{\rm sym}}^{{\rm cos}}\left(1\right)=-\frac{\omega_{\theta}}{\omega}\frac{\rho_{{\rm AHE}}}{\Delta\rho_{{\rm AMR}}}F_{{\rm asym}}^{{\rm cos}}\left(3\right),
\]
in the absence of the ISHE. Therefore, the factor above gives the
upper limit of the Lorentzian part due to the AHE. When $\omega_{{\rm \theta}}\left(\omega_{\phi}\right)$
possesses $\phi_{{\rm M}}$ dependence because of a magnetic anisotropy
field in-plane, Eq. (\ref{eq:Fsymasym}) should be recalculated by
considering the Fourier coefficients of $\omega_{\theta}$ and $\omega_{{\rm \phi}}$.
Note that when $\Delta\rho=0$ but $\rho_{{\rm AHE}}\neq0$, this
method can not be applied, and it is better to change a material for
the F layer or to try a measurement on the ferromagnetic layer thickness
dependence described in Sect. \ref{subsec:Thickness-dependence-of}.

\subsection{Frequency dependence }

Here, we focus on the difference in the frequency dependence of the
signals from the ISHE and the microwave effects. The frequency dependence
has attracted much attention for its nonlinear physics coming from
magnon-magnon interactions.\citep{kurebayashi2011controlled,castel2012frequency,harii2011frequency,sakimura2014nonlinear,chumak2015magnon}
Our interest is the linear excitation regime in which the derived
equations for the FMR can be used.\citep{iguchi2012spin} The heating
effects show a clear difference from the ISHE and rectification effects,
so that this method works effectively for removing the heating contribution.
\begin{figure}
\centering{}\includegraphics{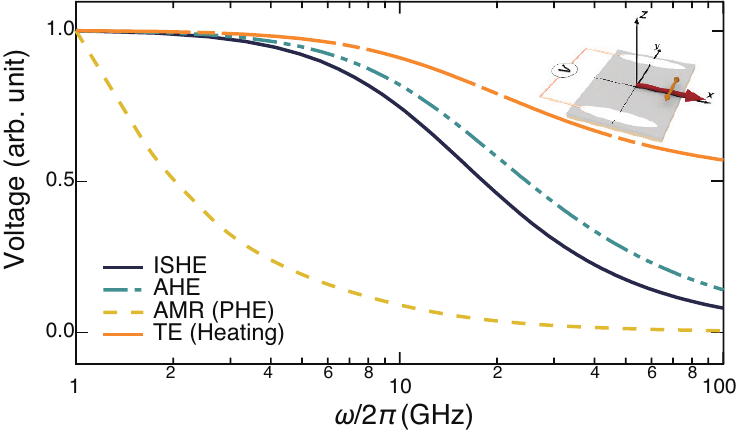}\caption{(Color online) Frequency dependence of signals induced by the ISHE,
rectification, and heating effects. The curves are normalized at $\omega/2\pi=1$
GHz. Other parameters are same as those used for the OP angular dependence
calculation.\label{fig:frequency-dependence}}
\end{figure}

In the calculation we consider a system with field along the $x$
and rf field along the $y$ axis, i.e. the IP excitation at $\theta_{{\rm M}}=90^{\circ}$
and $\phi_{{\rm M}}=0^{\circ}$. This configuration is often used
in measurements with a microstrip line or a coplanar waveguide. The
frequency dependence of the ISHE and rectification signals are respectively
given by 
\begin{equation}
E_{{\rm sym},{\rm ISHE}}^{y}=A_{{\rm ISHE}}\frac{\omega_{\phi}}{\left(\omega_{\theta}+\omega_{\phi}\right)^{2}}
\end{equation}
and 
\begin{equation}
E_{{\rm sym},{\rm rect}}^{y}=A{}_{{\rm rect}}\frac{\left(\omega\rho_{{\rm AHE}}{\rm Re}\left[j_{{\rm rf}}^{x*}\right]+\omega_{\phi}\Delta\rho_{{\rm AMR}}{\rm Im}\left[j_{{\rm rf}}^{x*}\right]\right)}{\omega\left(\omega_{\theta}+\omega_{\phi}\right)}.
\end{equation}
The electromotive force due to the heating, proportional to microwave
absorption $\Delta P$, is given by 
\begin{equation}
E_{{\rm sym},{\rm TE}}^{y}=A{}_{{\rm TE}}\frac{\omega_{\phi}}{\left(\omega_{\theta}+\omega_{\phi}\right)}
\end{equation}
with $A_{{\rm TE}}=\frac{\gamma\mu_{0}h_{{\rm rf}}^{2}}{2\alpha}\left(A_{{\rm TTE}}+A_{{\rm SE}\left({\rm MSSW}\right)}\right)$.
Figure \ref{fig:frequency-dependence} shows the calculated frequency
dependence of the ISHE, rectification effects, and heating effects.
Since the ISHE and AHE are similar, the separation between the ISHE
and rectification signals based on the frequency will not be so accurate.
The difference between the SP and the rectification effects comes
from the ellipticity of magnetization precession due to demagnetizing
and anisotropic fields. At frequencies $\omega>\gamma I_{{\rm eff}}$,the
large external field necessary for the FMR makes the precession trajectory
circular, so that both of them become proportional to $1/\omega$.
The heating is proportional to $\Delta P$ {[}Eq. (\ref{eq:delP}){]}
and is proportional to $\omega_{\theta\left(\phi\right)}/\omega$,
which reaches constant at high frequencies. This feature is distinct
from the others. The studies on the frequency dependence\citep{kurebayashi2011controlled,castel2012frequency,harii2011frequency,sakimura2014nonlinear,iguchi2012spin}
indicates that the thermoelectric contribution in the ferrimagnetic
insulator/Pt bilayer systems is not dominant for the microwave spin
pumping experiments.

\section{Summary\label{sec:Summary}}

\begin{figure}
\centering{}\includegraphics[width=12cm]{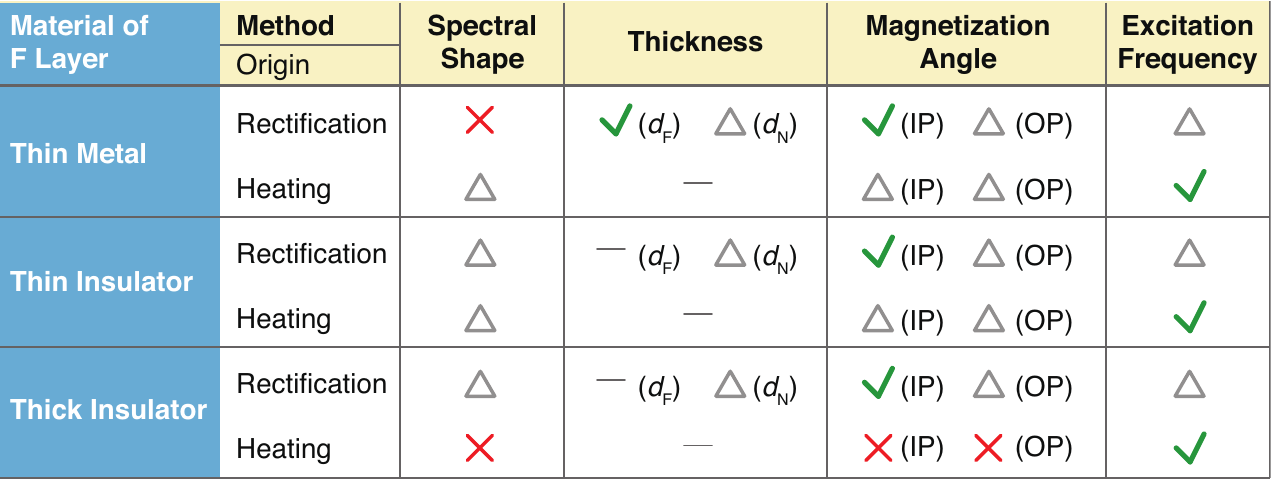}\caption{(Color online) Summary of methods for measuring the SP-induced ISHE.
It is written which effect can be separated (\Checkmark ) or cannot
be separated ($\times$) by using combination of the corresponding
material for the spin injector (F) layer and the measurement. $-$
denotes the non applicable separation method. $\triangle$ denotes
the moderate separation method when galvanomagnetic or transverse
thermoelectric effects are not negligibly small. Here, we assume that
the MSSW heating is dominant among the heating-induced signals. The
use of a thin ferrimagnetic insulator is the best way to explore spin
physics by using the spin pumping, because the magnitude of the galvanomagnetic
effects and MSSW heating is expected to be small compared to that
in metallic and thick ferromagnets. \label{fig:Summary-of-how}}
\end{figure}
In this article, we reviewed voltage generation by the SP-induced
ISHE, the rectification effects due to galvanomagnetic effects, and
the heating effects due to thermoelectric effects. The electric detection
of a spin current induced by the SP using the ISHE is a strong method
to study spin physics in a material of interest. The key for the study
is clear separation between the ISHE and the other extrinsic contributions.
In some configurations, they look similar to each other, but by employing
the separation methods introduced here, one can perform a reliable
measurement with high accuracy. Figure \ref{fig:Summary-of-how} summarizes
the recommended method by which the accuracy can be easily obtained.
For a better experiment, a material of interest should be on top of
a thin ferrimagnetic insulator film, which reduces both the rectification
and heating effects. For systems with metallic ferromagnet, the IP
magnetization angular dependence is the best configuration to clarify
the differences among the voltage signals of the ISHE and the rectification
effects, because the ISHE is sensitive to spin polarization while
the AMR is sensitive to the magnitude than the polarization. The separation
schemes discussed in this article provides a better way to extract
the SP-originated signals. We thus anticipate that the experimental
schemes help further material investigations and contribute to the
development of novel spintronic devices.

\section*{Acknowledgment}

The authors thank K. Sato, K. Uchida, and D. Hirobe for valuable discussions
and suggestions. This work was supported by Grant-in-Aid for Scientific
Research on Innovative Area, ``Nano Spin Conversion Science\textquotedblright{}
(No. 26103002, 26103005) from MEXT, Japan, and E-IMR, Tohoku University. 

\bibliographystyle{jpsj}
\bibliography{reviewref}

\end{document}